\documentclass[sigconf]{acmart}
\pdfoutput=1

\usepackage{algpseudocode}

\usepackage{algorithm}

\usepackage{enumitem}
\usepackage{bm}

\AtBeginDocument{%
  \providecommand\BibTeX{{%
    \normalfont B\kern-0.5em{\scshape i\kern-0.25em b}\kern-0.8em\TeX}}}

\setcopyright{iw3c2w3}
\copyrightyear{2021}
\acmYear{2021}
\acmDOI{10.1145/3442381.3449835}

\acmConference[WWW '21]{Proceedings of the Web Conference 2021}{April 19--23, 2021}{Ljubljana, Slovenia}
\acmBooktitle{Proceedings of the Web Conference 2021 (WWW '21), April 19--23, 2021,
Ljubljana, Slovenia}
\acmPrice{}
\acmISBN{978-1-4503-8312-7/21/04}

\begin{document}

\title{DGCN: Diversified Recommendation with Graph Convolutional Networks}


\author{Yu Zheng$^{1,2}$, Chen Gao$^{1,2}$, Liang Chen$^{3}$, Depeng Jin$^{1,2\dagger}$, Yong Li$^{1,2}$}
\thanks{$\dagger$Corresponding author.}
\affiliation{%
 \institution{$^1$Beijing National Research Center for Information Science and Technology}
 \institution{$^2$Department of Electronic Engineering, Tsinghua University}
 \institution{$^3$Sun Yat-sen University}}
\affiliation{%
  \institution{\{jindp, liyong07\}@tsinghua.edu.cn}
}
\renewcommand{\authors}{Yu Zheng, Chen Gao, Liang Chen, Depeng Jin, Yong Li}

\renewcommand{\shortauthors}{Zheng, et al.}

\begin{abstract}
  These years much effort has been devoted to improving the accuracy or relevance of the recommendation system. 
  Diversity, a crucial factor which measures the dissimilarity among the recommended items, received rather little scrutiny. 
  Directly related to user satisfaction, diversification is usually taken into consideration after generating the candidate items. 
  However, this decoupled design of diversification and candidate generation makes the whole system suboptimal. 
  In this paper, we aim at pushing the diversification to the upstream candidate generation stage, with the help of Graph Convolutional Networks (GCN). 
  Although GCN based recommendation algorithms have shown great power in modeling complex collaborative filtering effect to improve the accuracy of recommendation, how diversity changes is ignored in those advanced works. 
  We propose to perform rebalanced neighbor discovering, category-boosted negative sampling and adversarial learning on top of GCN. 
  We conduct extensive experiments on real-world datasets. 
  Experimental results verify the effectiveness of our proposed method on diversification. 
  Further ablation studies validate that our proposed method significantly alleviates the accuracy-diversity dilemma.
\end{abstract}

\ccsdesc[500]{Information systems~Collaborative filtering}

\keywords{Recommender systems, diversification, graph convolutional networks}


\maketitle

\section{Introduction} \label{sec::intro}

With the rapid development of the WEB, an intelligent algorithm called Recommendation System was proposed to overcome the information overflow problem \cite{cf}. 
The success of recommendation system has been verified in a number of scenarios including e-commerce \cite{amazon,he2016vbpr}, online news \cite{wu2019npa,wu2019neural} and multimedia contents \cite{chen2017attentive}. 
With the advancement of recommendation algorithms, much effort has been devoted to improving the accuracy of the recommended items. 
In other words, the accuracy serves as the dominant target or even the only target in most of the recent research. 
Chasing for higher accuracy, more attributes are incorporated \cite{FM,guo2017deepfm,he2017neural,lian2018xdeepfm}, and more complicated models are proposed \cite{youtube,din,airbnb,DLRM}.

However, an accurate recommendation is not necessarily a satisfactory one \cite{solving}. 
When users access the web, finding the exactly accurate contents is just one of their many needs. 
For example, users spend much time in browsing news-feed applications for discovering something novel \cite{kapoor2015like}. 
From the angle of user satisfaction, relevance is never the only rule of thumb. 
Many factors other than relevance, also influence how users perceive the recommennded contents, such as freshness, diversity, explainability and so on. 
Among those metrics affecting user satisfaction, \textbf{diversity} directly determines user engagement in recommendation scenarios \cite{wilhelm2018practical}. 
Specifically, not only the similarity between the user and the recommended items matters, but the dissimilarity among the recommended items reflects the recommendation effect as well. 
Without the diversity of the recommended items, users are likely to be exposed of repetitive items. 
That is to say, although the information overload problem is alleviated, another problem of information redundancy is brought in by recommendation system \cite{wilhelm2018practical}.

In order to guarantee user satisfaction, three directions of approaches, namely post-processing, Determinantal Point Process (DPP) and Learning To Rank (LTR), have been proposed to improve the diversity of the recommended results \cite{recent}. 
In the early stage of diversified recommendation, a re-rank or post-processing module is appended after the generation of recommended candidates. 
The order of the items is determined by heuristics to balance between relevance and diversity. 
A bunch of solutions in this research line were proposed \cite{mmr,topic,msd,dum,direc,pmf_alpha_beta,entropy}. 
Independent with candidate generation, the re-rank strategy is decoupled from the optimization of the recommendation model. 
Thus the diversification signal is not reflected in upstream relevance matching models, which increases the risk of the final recommendation being suboptimal. 
Recently, another direction of research takes advantage of DPP \cite{gillenwater2014expectation,chen2018fast,gartrell2017low,gillenwater2019tree,warlop2019tensorized,wilhelm2018practical} to replace the heuristics in post-processing based mothods, but the diversification process of DPP is still conducted after the 
 generation stage. 
 To address this problem, a series of methods based on LTR \cite{dcf,li2017learning} were proposed which target on directly generating an ordered list of items rather than a candidate set. 
 However, it is tricky to construct an appropriate listwise dataset for such methods. 
 To summarize, existing solutions based on post-processing or DPP aim to improve the diversity leaving the matching process untouched. 
 However, the overall performance of a recommendation system greatly depends on the representations of users and items learned in the matching process. 
 Thus it remains uncertain whether the decoupled design diversifies the recommendation with acceptable loss in accuracy. 
 In terms of LTR based methods, practical issues exist because of the difficulty in collecting feasible datasets.

Since the interactions between users and items can be naturally represented as a heterogeneous graph (a bipartite of users and items), a number of graph based recommendation algorithms were proposed which either utilize rather simple random walk \cite{pixie,jiang2018recommendation} or more complicated methods like GCN \cite{pinsage,NGCF,KGAT,wang2019knowledge,wu2019session}. 
In terms of the user-item bipartite graph, higher order neighbors of a user tend to cover more diverse items, because these neighbors contain not only the given user's interacted items, but also other similar users' preferred items. 
Therefore, it is advantageous to perform diversification on a graph, since the high order connectivity makes diverse items more reachable. 
Furthermore, performing diversification in GCN also alleviates the aforementioned problem of existing works which separate diversification from the upstream relevance matching model.
However, without specific designs for diversity, those high order connections might not be automatically utilized to find items which are not similar to each other. 
For example, the recommendation system can easily learn to provide items of the most interacted categories because they take up the majority of the edges on the graph. 
Nevertheless, those GCN based algorithms mainly focus on improving the accuracy, while ignoring how diversity is impacted by the much more complicated GCN model.

In our work, we focus on category diversification in recommendation with the help of GCN. 
We develop rebalanced neighbor discovering for GCN to make items of disadvantaged categories more reachable. 
We make adjustments to the negative sampling process to boost the probability of sampling \textit{similar but negative} items. 
Furthermore, we employ adversarial learning to distill the implicit category preference in the learned embedding space. 
Through these special designs, we push the diversification process upstream into the matching stage and propose an end-to-end model called \textbf{D}iversified recommendation with \textbf{G}raph \textbf{C}onvolutional \textbf{N}etworks (DGCN). The main contributions of this paper are three-fold:
\begin{itemize}[leftmargin=*]
    \item We analyze the effect of existing diversification algorithms and propose a novel method to combine diversification with matching. The integrity of our method overcomes the problem of decoupled structure in existing works.
    \item Aiming to generate diverse and relevant items, we carefully design rebalanced neighbor discovering, category-boosted negative sampling and adversarial learning for GCN. An end-to-end model is developed for diversified recommendation.
    \item We utilize real-world datasets to evaluate the effectiveness of our proposed method. Experimental results demonstrate that diversity of recommendation is validly improved by our method. Furthermore, we conduct ablation studies to confirm the importance of each proposed component.
\end{itemize}

The remainder of the paper is organized as follows. First, we introduce a few preliminaries in Section \ref{sec::pre}. Then we elaborate our design of DGCN in Section \ref{sec::method} and conduct experiments in Section \ref{sec::exp}. After reviewing related work in Section \ref{sec::related}, we make some discussions and conclude the paper in Section \ref{sec::con}.
\section{Preliminaries} \label{sec::pre}

\subsection{Diversity}

As a matter of fact, diversity of recommendation can be either intra-user level or inter-user level \cite{chen2016personality}. 
Intra-user level diversity measures the dissimilarity of the recommended items of an individual user, while inter-user level focuses on the provided contents for different users. 
In this paper, we target on improving intra-user level diversity\footnote{in this paper, \textit{diversity} refers to \textit{intra-user level diversity} for simplicity} as most of the related research, and leave inter-user level diverisity (also known as \textit{decentration} or long-tail recommendation) for future work.

Diversity is often mixed up with serendipity or novelty. 
For example, suppose 70\% of the purchased items for a user are electronic devices, 20\% are clothes and the other 10\% are drinks. 
Then a recommended list of 10 items including five or more electronic devices, one or two clothes and one or two drinks is a much more diverse result than recommending ten electronic devices. 
Moreover, even though the user did not purchase any books in her interaction history, she might still have interests in reading, and serendipity stands for the capability of the recommendation system to provide items appealing to users but not realized by themselves (books for the user in this case).

In this paper, we focus on category diversification \cite{topic}. When users browse the recommended items, it is not user-friendly to provide a large amount of items belonging to the same category. We utilize three widely adopted metrics for diversity in our experiments:
\begin{itemize}[leftmargin=*]
    \item{\textbf{coverage}}: this metric measures the number of recommended categories. Coverage reflects the holistic and overall diversity of a recommendation system.
    \item{\textbf{entropy}}: this one focuses on the distribution on different categories. Using the previous example, the entropy value of four electronic devices, three clothes and three drinks is higher than recommending seven electronic devices and three drinks.
    \item{\textbf{gini index}}: this index is popularly adopted in economics to measure the wealth or income inequality, and it was further adapted and introduced to recommendation by \cite{antikacioglu2017post}. The number of items belonging to a specific category can be explained as the wealth of that category.
\end{itemize}
Note that in terms of coverage and entropy, higher value means stronger diversity, while for gini index it is the opposite (\textbf{lower is better}).

\subsection{Recommendation Pipeline}

\begin{figure}[t]
    \centering
    \includegraphics[width=\linewidth]{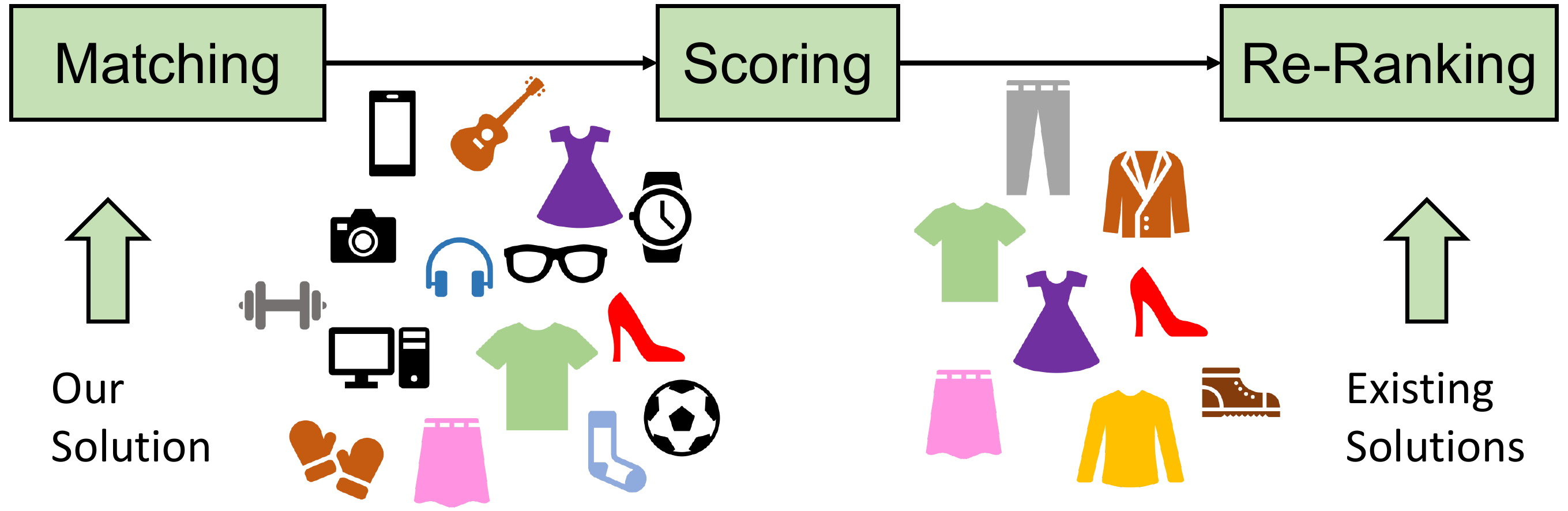}
    \caption{A typical recommendation pipeline.}
    \vspace{-10px}
    \label{fig::serving}
\end{figure}

\begin{figure}[t]
    \centering
    \includegraphics[width=\linewidth]{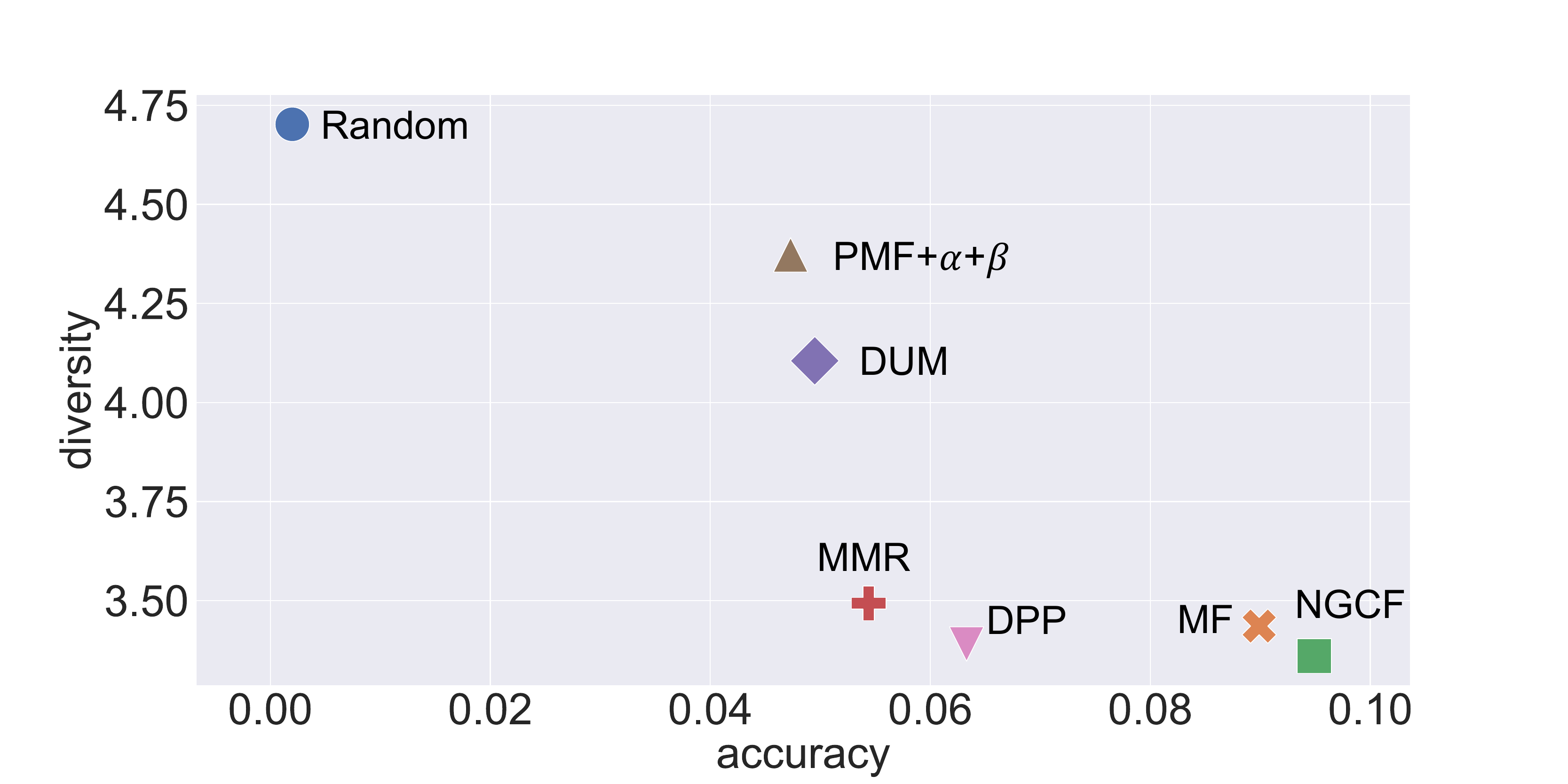}
    \vspace{-20px}
    \caption{The accuracy-diversity dilemma.}
    \vspace{-10px}
    \label{fig::dilemma}
\end{figure}

As illustrated in Figure \ref{fig::serving}, a typical pipeline for a recommender system is composed of three stages: (1) matching (candidate generation), (2) scoring, and (3) re-ranking. Several hundreds of items are selected in matching stage from a large item pool. Then, usually complicated deep learning models are adopted in scoring stage to estimate interaction probability and the top dozens of items are selected. In re-ranking, selected items are re-arranged to satisfy additional constraints.

With respect to diversity, it is widely adopted to impose heuristic rules in the re-ranking stage to diversify recommended items. 
Different re-ranking methods were proposed to balance between relevance and diversity \cite{mmr, msd, dum, wilhelm2018practical}. 
However, diversification in re-ranking is independent with upstream matching and scoring models, which makes the whole system suboptimal. 
Furthermore, with matching models unaware of diversification signals, the retrieved items from matching can be already redundant, which limits diversity from the source. 
In this paper, we aim at pushing diversification upwards. 
Specifically, we take diversity into consideration during matching, and propose an end-to-end method to provide diverse recommendation.

\subsection{Accuracy-Diversity dilemma}

Generally, when considering diversity, it is not easy to get rid of the so called accuracy-diversity dilemma \cite{solving}, especially in offline evaluations. 
That is, higher accuracy often means losing diversity to some extent. 
We compared several recommendation algorithms (random, matrix factorization \cite{MF} and neural graph collaborative filtering \cite{NGCF}), as well as a bunch of diversification algorithms (MMR \cite{mmr}, DUM \cite{dum}, PMF+$\alpha$+$\beta$ and DPP \cite{dppy}), utilizing a real world e-commerce dataset collected from Taobao\footnote{www.taobao.com}, which is the largest e-commerce platform in China. 
In order to verify the tradeoff between the two metrics, we plot the results of these methods in Figure \ref{fig::dilemma} with accuracy and diversity as the two axes.

From Figure \ref{fig::dilemma} we can observe that, with the recommendation algorithm getting more complicated, though more relevant products are provided, less categories are presented to customers. 
After introducing these diversification strategies, the diversity indeed gets promoted, while the accuracy is not guaranteed. 
Although there exist certain hyper-parameters in these diversification methods to balance the bias, later experiments show that it is rather difficult to find a satisfactory point.

\section{method} \label{sec::method}

\subsection{Overview}

\begin{figure*}[h]
    \centering
    \includegraphics[width=\linewidth]{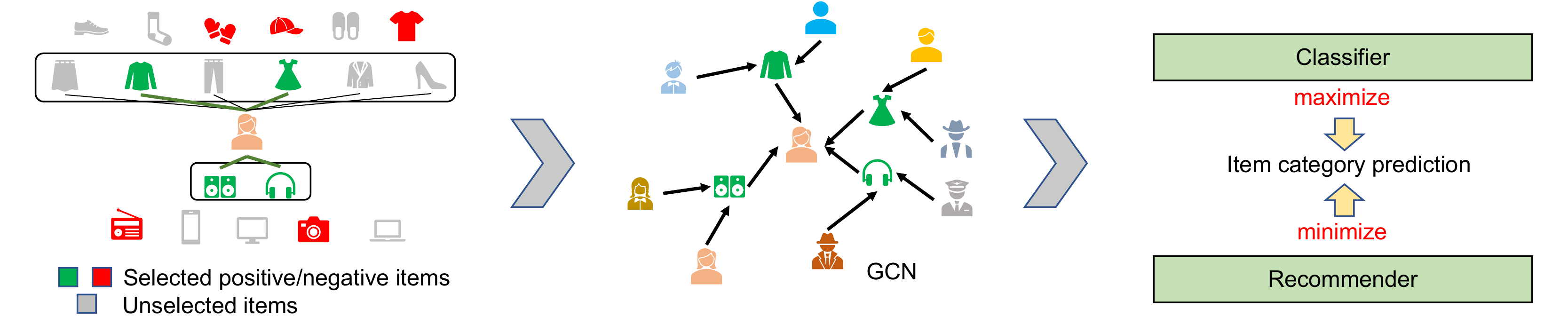}
    \caption{Overview of our proposed DGCN.}
    \vspace{-10px}
    \label{fig::pipeline}
\end{figure*}

As introduced previously, we incorporate diversification into the matching process with GCN. We aim at providing items correlated with users' interests while dissimilar with each other (diversity). Figure \ref{fig::pipeline} illustrates the holistic structure of our proposed DGCN.

In real-world recommendation scenarios, the interactive behaviors between users and items are always the strongest signals with respect to user preference modeling. Thus we first construct a graph to represent the interactions, which consists of two types of nodes (users and items), and edges between them stand for those behaviors. It is worthwhile to state that edges are undirected in our constructed bipartite graph. We propose rebalanced neighbor discovering to solve the problem of inconsistency between different categories. By applying GCN on top of the the sampled sub-graph, node features propagate back and forth between users and items, which accurately simulates the collaborative filtering effect. With the goal of making diverse categories more accessible, we make adjustments to the negative sampling process, boosting the probability of selecting similar items. Furthermore, we add an adversarial task of item category classification to strengthen the diversity.

Our proposed method is featured with the following three special designs targeting on diversification: (1) Rebalanced Neighbor Discovering, (2) Category-Boosted Negative Sampling and (3) Adversarial Learning.
\begin{itemize}[leftmargin=*]
    \item{\textbf{Rebalanced Neighbor Discovering}} To discover more diverse items on the graph, we design a neighbor sampler based on the distribution of the neighbors. We increase the probability of selecting items of disadvantaged categories, and limit the effect of those dominant categories. With the guidance of the neighbor sampler, items of multiple categories are more reachable. 
    \item{\textbf{Category-Boosted Negative Sampling}} Unlike random negative sampling, we propose to choose those \textit{similar but negative} items with a boosted higher probability. By distorting the distribution of negative samples, representations for users and items are learned in a finer level, where the recommendation system can determinate the user preferences among similar items.
    \item{\textbf{Adversarial Learning}} We leverage the technique of adversarial learning, playing a min-max game on item category classification. With an extra adversarial task, we distill users' category preferences from item preferences, which makes the learned embeddings category-free. And consequently, neighbors in this embedding space will cover more categories.
\end{itemize}

In the following sections, we first introduce the architecture of the adopted GCN, and then we elaborate on the three special designs for diversity one after another.

\subsection{GCN}

Our GCN is composed of an embedding layer and a stack of graph convolutional layers, where each graph convolutional layer contains a broadcast operation and an aggregation operation.

\subsubsection{Embedding Layer}

Inspired by representations for words and phrases \cite{word2vec}, the embedding technique has been successfully employed in recommendation system \cite{airbnb}. In our work, the inputs of GCN are simply ID features for users and items. Following the widely adopted embedding strategy, we transform the one-hot ID feature to a dense vector, thus we have the following embedding look-up table:
\begin{equation}
    \bm{E} = [\bm{e_{u1}}, \dots , \bm{e_{uM}}, \bm{e_{i1}}, \dots ,\bm{e_{iN}}],
\end{equation}
where $M$ is the number of users, and $N$ is the numberof items. We represent each node with a separate embedding $\bm{e} \in \mathbb{R}^d$, in which $d$ is the embedding size. It is worthwhile to note that the embeddings are learnable parameters and further fed into GCN for message passing on the graph. Thus the embedding can be regarded as the node feature at layer 0, i.e. $\bm{h_v^0}$ which will be introduced later.

\subsubsection{Graph Convolutional Layer}

We perform embedding broadcast and aggregation in the graph convolutional layer. In other words, within a graph convolutional layer, each node broadcasts its current embedding to all its neighbors and aggregates all the messages sent to it to update its embedding. In terms of neighbor aggregation, we utilize average pooling combined with a feature transform matrix and a nonlinear activation function. Formally, we denote the feature vector of node $v$ at the $k$-th layer as $\bm{h_v^k}$, and the update rules are illustrated as follows:
\begin{equation}
\begin{aligned}
    \bm{h_{AGG}^k} = & MEAN({\bm{h_j^{k-1}}, \forall j \in \mathcal{N}(v)}), \\
    \bm{h_v^k} = & \tanh(\mathbf{W^k}\bm{h_{AGG}^k})
\end{aligned}
\end{equation}
where $\mathcal{N}(v)$ represents the set for sampled neighbors of node $v$. As investigated in \cite{sgc}, adding self loops is of crucial importance in graph convolutional networks, since it compresses the spectrum of the normalized Laplacian. Therefore, we also insert node $v$ itself into $\mathcal{N}(v)$. In this way, node embeddings propagate on the graph in a layer-wise manner.

\subsubsection{Interaction Modeling} 

With respect to interaction modeling in the matching stage, heavy computation such as the inference computing of neural networks is impractical due to the large item pool and the strict latency requirements. For embedding based matching model, inner product and L2 distance are widely used. Furthermore, at online serving time, these simple rather effective interaction modeling methods can be greatly accelerated with the help of nearest neighbor search algorithms. Therefore, we use the representations of users and items at the last graph convolutional layer and take inner product of them to estimate the interaction probability:
\begin{equation}
    p_{u,i} = <\bm{h_u^{K}}, \bm{h_i^{K}}>,
\end{equation}
where $K$ is the depth of the graph neural networks. During evaluations, top items with respect to $p_{u,i}$ are selected as recommended items for a given user $u$.

\subsubsection{Prevent Overfitting}
To prevent our model from overfitting, we perform dropout \cite{dropout} on the feature level. To be specific, we randomly drop the intermediate node embeddings between consecutive graph convolutional layers with probability $p$, where $p$ is a hyper-parameter in our method.

With great capability in learning representations for graph structures, GCN has been shown effective to capture the collaborative effect on the user-item bipartite graph, which improves the accuracy significantly. However, utilizing the high order connectivity for diversification has received little scrutiny. We then introduce our special designs for diversity in the proposed DGCN.

\subsection{Rebalanced Neighbor Discovering}

In the matching stage, usually the recommendation system retrieves items from a large corpus which is of million-scale or even billion-scale \cite{pinsage}. Feeding the whole graph consisting of millions of nodes to a GCN suffers from highly inefficient computation and heavy resources usage. Moreover, it is difficult to implement mini-batch training on the whole graph. Thus a neighbor sampler is often employed to sample a sub-graph from the original large one for efficient training \cite{graphsage}. With the help of the neighbor sampler, inductive learning on the graph is accomplished and it has been proved scalable to billion-scale graphs \cite{pinsage}. Specifically, the neighbor sampler generates a \textit{Node Flow}, which is a sub-graph with multiple layers, where edges only exist in consecutive layers.

\begin{figure}[h]
    \centering
    \includegraphics[width=\linewidth]{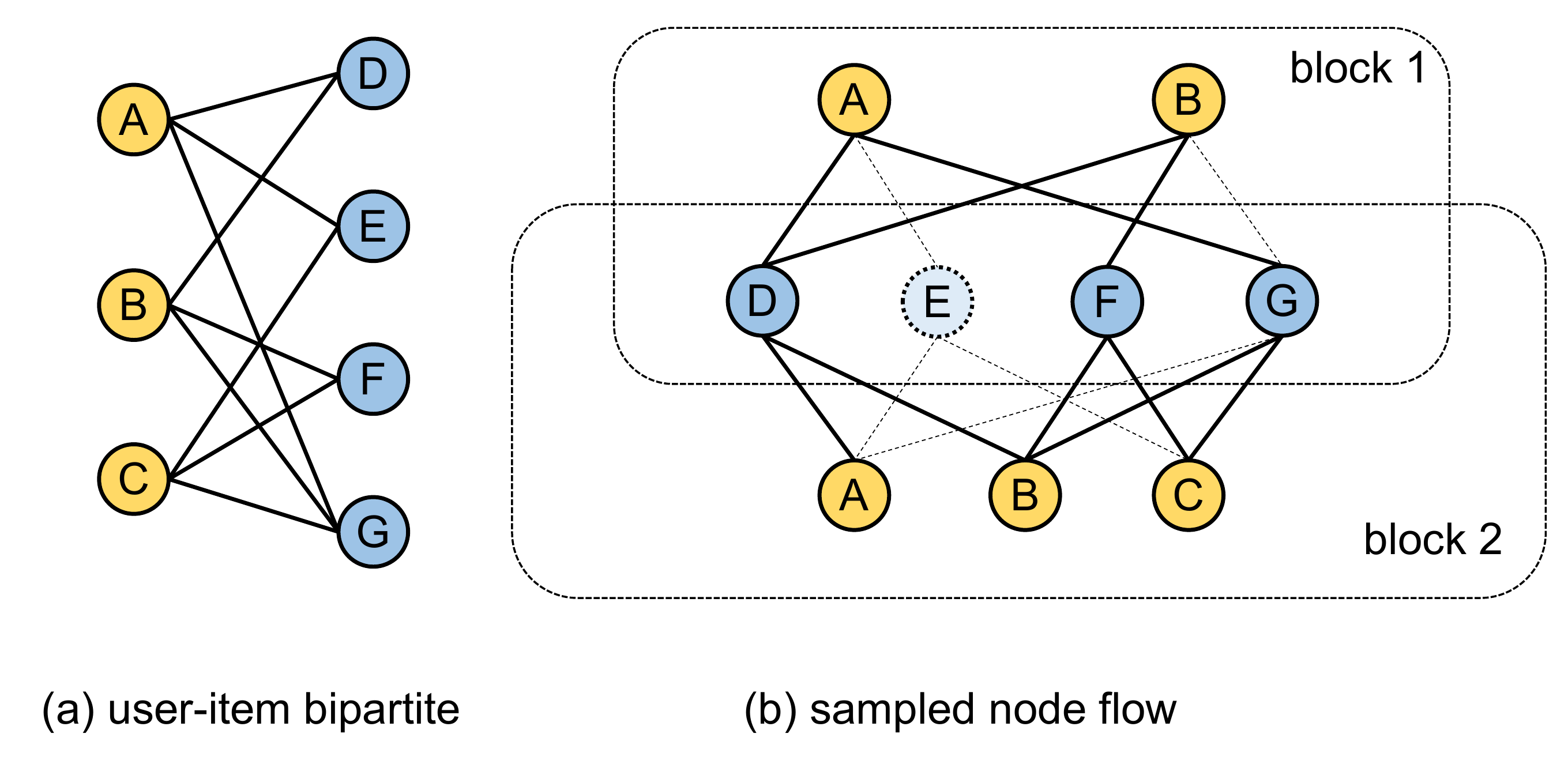}
    \vspace{-20px}
    \caption{Node Flow generated by the neighbor sampler for a 2-layer GCN. In this example, node A and node B are the initial seed nodes. For each node, the neighbor sampler randomly samples two neighbors. The first GCN layer performs convolution in block 1, and all the activated neighbors in block 1 form the seed nodes for sampling block 2, which corresponds to the second GCN layer.}
    \vspace{-10px}
    \label{fig::nodeflow}
\end{figure}
Figure \ref{fig::nodeflow} serves as a toy example for neighbor discovering. Specifically, during the training process, a mini-batch consists of a certain number (i.e. batchsize) of users and items, and these user nodes and item nodes in a single batch form the set of seed nodes. The neighbor sampler randomly selects their neighbors and extracts the sub-graph. For GCN deeper than one layer, this neighbor discovering process will repeat recursively, which means the sampled neighbors become seed nodes for the next hop. It is clearly illustrated in Figure \ref{fig::nodeflow} that edges exist in consecutive layers and connected layers form a block. Graph convolutional operations are performed block by block, where each graph convolutional layer corresponds with one block.

However, the above neighbor discovering strategy leaves aside the problem of diversity. In real-world recommendation scenarios, items of distinct categories are supposed to be regarded differently because users perceive them differently. In other words, there exist dominant categories and disadvantaged categories according to users' interaction history, where users engage more with items of dominant categories and spend much less time in disadvantaged categories. A diversified recommendation system is capable of providing items from not only dominant categories but also disadvantaged categories. From the view of the graph, dominant categories of a user become the mainstream in the message flow on the graph, because they take up most of the in-edges to the user node. Thus, without distinguishing between dominant categories and disadvantaged categories in neighbor discovering, the learned user embeddings are likely to be too close to items embeddings of dominant categories, which makes it rather difficult to retrieve items from other categories, and thus limits the diveristy of recommendation.

In our work, we make adjustments to the neighbor discovering process, with special emphasis on category diversification. Specifically, we boost the probability of sampling items from disadvantaged categories and restrict the number of selected items from dominant categories. The rebalanced neighbor discovering algorithm is illustrated in Algorithm \ref{alg::neighbor} and \ref{alg::rebalance}.
Due to space limit, we omit the descriptions for \verb|GetNeighbors| and \verb|SampleWithProbability| in Algorithm \ref{alg::neighbor}, which are simple look-up operations in the adjacency list and random choice with a  given distribution. For a user node, we first conduct histogram of item categories to find dominant ones and disadvantaged ones. By taking the inverse of the histogram, we boost the probability of sampling disadvantaged categories and lower the priority for dominant categories. A rebalance weight $\alpha$ is introduced to control the bias. For an item node, we equally treat all the user nodes linked to it and sample its neighbors uniformly. Through rebalanced neighbor discovering, items of more categories are selected, which in turn makes the user embeddings absorb more diverse item embeddings according to the logic of GCN. Therefore, items of superior diversity are recommended by retrieving items from the embedding space with the learned user embedding as the query vector.
\begin{algorithm}[t]
    \caption{Rebalanced Neighbor Discovering}
    \begin{flushleft}
        \textbf{INPUT:} Graph $\mathcal{G}(\mathcal{V}, \mathcal{E})$; GCN depth $K$; seed nodes set $\mathcal{S}$; number of neighbors to sample $n$; item-category table $\mathcal{C}$; rebalance weight $\alpha$ \\
        \textbf{OUTPUT:} Node Flow $\mathcal{L}$ \\
    \end{flushleft}
    \label{alg::neighbor}
    \begin{algorithmic}[1]
        \State {$\mathcal{L} \leftarrow \mathrm{EmptyList}()$}
        \For {$k \leftarrow 1 \, to \, K$}
            \State {$\mathcal{L}^k \leftarrow \mathrm{EmptySet}()$}
            \ForAll {node $v \in \mathcal{S}$}
                \State {$\mathcal{N}(v) \leftarrow \mathrm{GetNeighbors}(\mathcal{G}, v)$}
                \If {$v$ is a user node}
                    \State {$p \leftarrow \mathrm{HistogramAndRebalance}(\mathcal{N}(v),\mathcal{C}, \alpha)$}
                \ElsIf {$v$ is an item node}
                    \State {$p \leftarrow \mathrm{UniformDistribution}()$}
                \EndIf
                \State {$\mathcal{L}_v^k \leftarrow \mathrm{SampleWithProbability}(\mathcal{N}(v), n, p)$}
                \State {$\mathcal{L}^k \leftarrow \mathrm{Union}(\mathcal{L}^k,  \mathcal{L}_v^k)$}
            \EndFor
            \State {append set $\mathcal{L}^k$ to $\mathcal{L}$}
        \EndFor
        \State {\Return $\mathcal{L}$}
    \end{algorithmic}
\end{algorithm}

\begin{algorithm}[t]
    \caption{HistogramAndRebalance}
    \begin{flushleft}
        \textbf{INPUT:} User node $u$'s neighbors $\mathcal{N}(u)$; item-category table $\mathcal{C}$; rebalance weight $\alpha$\\
        \textbf{OUTPUT:} Sample probability over node $v$'s neighbors $p$ \\
    \end{flushleft}
    \label{alg::rebalance}
    \begin{algorithmic}[1]
        \State {$H \leftarrow \mathrm{ComputeCategoryHistogram}(\mathcal{N}(v), \mathcal{C})$}
        \ForAll {node $i \in \mathcal{N}(u)$}
            \State {$p(i) \leftarrow 1/H(\mathcal{C}(i))$}
            \State {$p(i) \leftarrow p(i)^{\alpha}$}
        \EndFor
        \State {$p \leftarrow \mathrm{Normalize}(p)$}
        \State {\Return $p$}
    \end{algorithmic}
\end{algorithm}

\subsection{Category-Boosted Negative Sampling}

One of the main challenges for matching is the so called implicit feedback \cite{rendle2009bpr}. That is, only positive samples are accessible to the recommendation system while negative samples are inferred from the uninteracted items. This implicit protocol means negative samples are not necessarily the ones users truly dislike. In practice, negative instances are generated by randomly sampling from those uninteracted items. When training recommendation models, each positive sample is paired with a certain number (i.e. \textit{negative sample rate}) of negative samples. By optimizing with pointwise \cite{ncf} or pairwise \cite{rendle2009bpr} loss function, positive item embeddings are learned to be close to user embeddings, while negative item embeddings are pushed off to the opposite direction.

Several works were proposed to improve the design of the negative sampler \cite{ding2018www,ding2019reinforced}, aiming at promoting the recommendation accuracy. Nevertheless, few works investigate the potential of negative sampling in diversification. In our work, we propose to choose those \textit{similar but negative} items, which means items of the same category with the positive sample. By sampling negative items from the \textit{positive category}, the recommendation model is optimized to distinguish users' preference within a category. And those negative items in the same category are less likely to be retrieved, which increases the possibility of recommending items from other more diverse categories. The negative sampling strategy is explained in Algorithm \ref{alg::negative}.

\begin{algorithm}[t]
    \caption{Category-Boosted Negative Sampling}
    \begin{flushleft}
        \textbf{INPUT:} Positive samples $\boldsymbol{P}$; item set $\boldsymbol{I}$; negative sample rate $T$; item-category table $\mathcal{C}$; similar sampling weight $\beta$\\
        \textbf{OUTPUT:} Training samples $\boldsymbol{\Omega}$ \\
    \end{flushleft}
    \label{alg::negative}
    \begin{algorithmic}[1]
        \State {$\boldsymbol{\Omega} \leftarrow \boldsymbol{P}$}
        \ForAll {positive sample $(u, i, \mathrm{True}) \in \boldsymbol{P}$}
            \State {$\boldsymbol{N} \leftarrow \boldsymbol{I} \setminus i$}
            \State {$\boldsymbol{S} \leftarrow \boldsymbol{I}_{\mathcal{C}(i)} \setminus i$}
            \For {$t \leftarrow 1 \, to \, T$}
                \State {$r \leftarrow \mathrm{RandomFloat(0,1)}$}
                \If {$r < \beta$}
                    \State {$i_t \leftarrow \mathrm{Sample}(\boldsymbol{S})$}
                \Else
                    \State {$i_t \leftarrow \mathrm{Sample}(\boldsymbol{N})$}
                \EndIf
                \State {$\boldsymbol{\Omega} \leftarrow \boldsymbol{\Omega} + (u, i_t, \mathrm{False})$}
            \EndFor
        \EndFor
        \State {\Return $\boldsymbol{\Omega}$}
    \end{algorithmic}
\end{algorithm}

\begin{figure}[t]
    \centering
    \includegraphics[width=\linewidth]{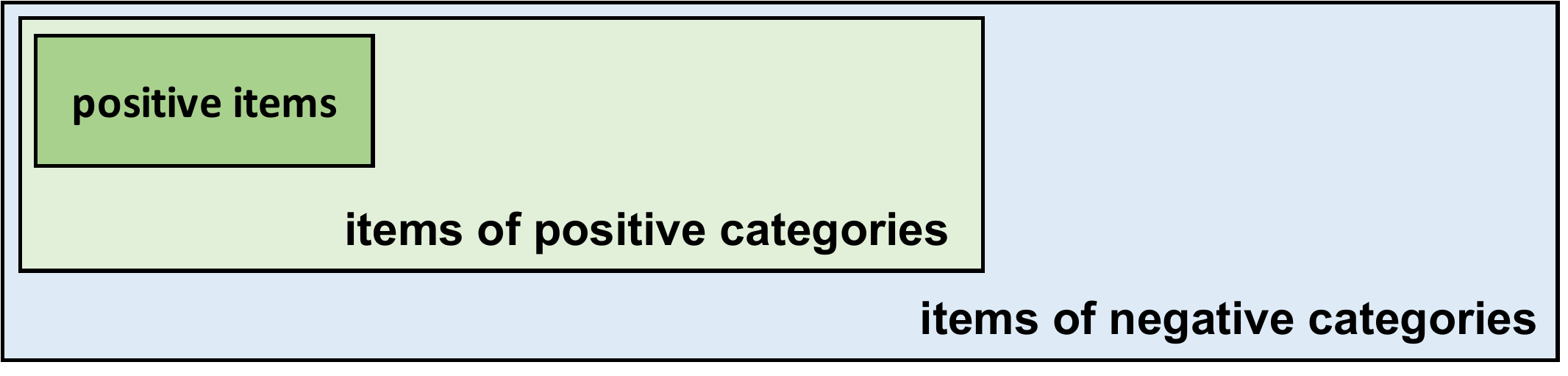}
    \caption{An illustration of the sample space. Negative instances are sampled from outside positive items. We propose category-boosted negative sampling which boosts the probability of sampling from items of positive categories (the light green area).}
    \vspace{-10px}
    \label{fig::samplespace}
\end{figure}

A hyper-parameter $\beta$ is introduced to manage the proportion of samples from similar items. With more \textit{similar but negative} items in training samples, the representations of user and items are learned in a finer level, which empowers the recommendation system to capture users' interests from more diverse categories. As illustrated in Figure \ref{fig::samplespace}, items from positive categories are sampled more as negative instances, which increases the possibility for positive items from negative categories to be recommended and thus more diverse candidates are generated.

\subsection{Adversarial Learning}

With respect to model training, most recommendation models are trained with a single target concerning accuracy. Though the multi-task framework has been applied in recommendation for multi-behavior modeling \cite{gao2018learning}, relevance of the results still served as the core goal and diversity of recommendation was ignored. With only one optimization object of accuracy, users' category preference is implicitly learned from users' item preference. Taking the same example utilized in Section \ref{sec::pre}, the recommendation system might learn the user's interests on the whole category (i.e. electronic devices), while fails to distinguish between the user's specific preference on different electronic devices.

Without distillation of the implicit category preference captured in the recommendation model, more items of the positive categories will be recommended, which limits the chance for more diverse items to be exposed to users. Inspired by the progress made in generative models \cite{gan,dcgan}, we propose to add an extra adversarial task of item category classification to achieve the goal of distillation and further enhance the diversity. Specifically, we augment the recommendation model with a classifier based on the learned item embeddings. We hope the classifier to predict the category of the item from the item embedding as accurate as possible, and expect the recommendation model to generate item embeddings which best \textit{fool} the classifier.

In our experiments, we adopt a fully connected layer as the classifier and use cross entropy loss for optimization. With respect to recommendation, we use log loss \cite{ncf} which is shown effective in experiments. Take a single training sample $(u, i, y, c)$ as an example, where $y$ is either 0 or 1 represents the user-item interaction groundtruth and $c$ is item $i$'s corresponding category. The loss function for recommendation is formulated as follows:
\begin{equation}
    \begin{aligned}
        \hat{y} =& <\bm{h_u^K}, \bm{h_i^K}>, \\
        L_r(u,i,y) =& -[y \cdot \log{\sigma(\hat{y})} + (1 - y) \cdot \log{\sigma(1 - \hat{y})}],
    \end{aligned}
\end{equation}
where $h_u^K$ and $h_i^K$ are the representations learned by GCN (at the last layer). The loss function for item category classification is:
\begin{equation}
    \begin{aligned}
        \hat{c} =& \bm{Wh_i^K} \\
        L_c(i,c) =& -\hat{c}[c] + \log{(\sum_jexp(\hat{c}[j]))}
    \end{aligned}
\end{equation}
Under the setting of adversarial learning, the object for the item category classifier is to minimize $L_c$, and the object for the recommendation model is to minimize $L_r - \gamma L_c$, where $\gamma$ is introduced to balance the main task and the additional adversarial task.

With respect to the classifier, the classification loss is minimized by finding clusters of item embeddings. While for the recommendation model, the classification loss is reversed which pushes item embeddings of the same category far from each other and not to form clusters. Meanwhile, the main task of minimizing the recommendation loss forces the learned embedding space to retain user preference semantics.

\begin{figure}[t]
    \centering
    \includegraphics[width=\linewidth]{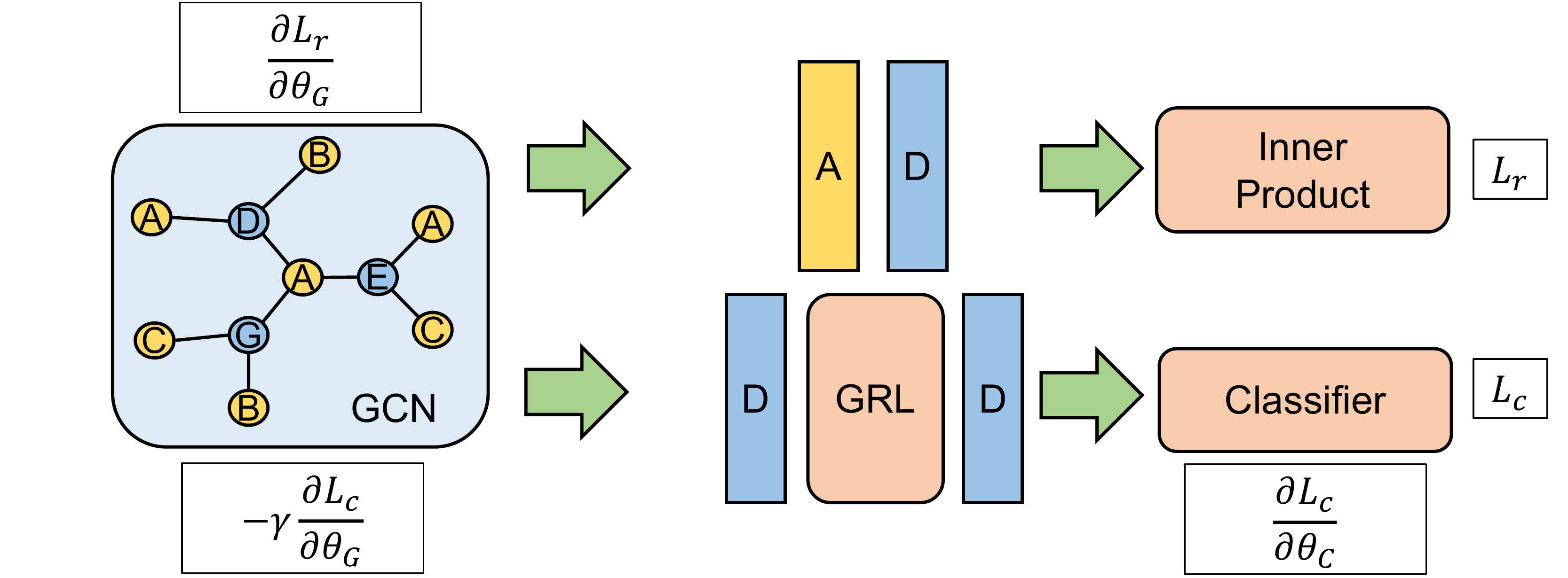}
    \caption{Implementaion of adversarial learning. Gradients are shown in boxes.}
    \vspace{-10px}
    \label{fig::grl}
\end{figure}

In terms of implementation, adversarial learning can be elegantly accomplished by inserting a Gradient Reversal Layer (GRL) in the middle of the back propagation process, which was first introduced in Domain Adaptation Networks (DAN) \cite{DAN}. We adopt this strategy in our work.
Using the same notations of the previous section, we expect the classifier to minimize $L_c$, while force the GCN to maximize $L_c$. As illustrated in Figure \ref{fig::grl}, we insert a GRL in between of the learned item embeddings from GCN and the fully connected classifier.
During the back propagation process, the gradients for minimizing the classification loss flow backward through the classifier, and after the GRL, the gradients will be reversed, which further flow to GCN. That is, we perform gradient descent on the parameters of the classifier, while perform gradient ascent on the parameters of GCN, with respect to $L_c$. For $L_r$, gradient descent is applied to GCN. Through this subtle design, we successfully implement the adversarial learning task. 

With the help of adversarial learning, the learned representations of users and items to great extent reserve the item-level interests while squeeze out the category-level interests. Therefore, positive items from negative categories are drawn near to users, and negative items from positive categories are pushed away. Consequently, neighbors in the embedding space will cover items of more diverse categories.

\section{Experiments} \label{sec::exp}

In this section, we conduct experiments to answer the following research questions:

\begin{itemize}[leftmargin=*]
    \item \textbf{RQ1}: How does the proposed method perform compared with other diversified recommendation algorithms?
    \item \textbf{RQ2}: What is the effect of each proposed component in DGCN?
    \item \textbf{RQ3}: How to perform trade-off between accuracy and diversity using DGCN?
\end{itemize}

\subsection{Experimental Settings}

\subsubsection{Datasets and Evaluation Protocols}

To evaluate the performance of our proposed method, we utilize three real-world datasets: Taobao, Beibei and Million Song Dataset (MSD). The three datasets vary in scale and density. Basic statistics of the datasets are summarized in Table \ref{tab::dataset}.

\begin{itemize}[leftmargin=*]
    \item{\textbf{Taobao}}: This dataset \cite{zhu2018learning,zhu2019joint} contains the behaviors of users on taobao.com including click, purchase, adding item to shopping cart and item favoring during November 25 to December 03, 2017, which was provided by Alimama\footnote{https://tianchi.aliyun.com/dataset/dataDetail?dataId=649}. We regard all the aforementioned behaviors as positive samples and randomly select about 10\% users with uniform probability. We adopt 10-core settings which means only retaining users and items with at least 10 interactions.
    \item{\textbf{Beibei}}: This dataset \cite{gao2018learning} is collected from one of the largest e-commerce platforms\footnote{https://www.beibei.com} in China which records the purchase behaviors during July 1 to July 31, 2017. We also utilize the 10-core settings to guarantee the data quality.
    \item{\textbf{MSD}}: This dataset \cite{Bertin-Mahieux2011} contains the listening history for 1M users, and is has been utilized to evaluate diversified music recommendation algorithms \cite{chen2018fast}. We extract a subset of the dataset and use 10-core settings to filter out inactive entities.
\end{itemize}

\begin{table}
    \caption{Statistics of the datasets.}
    \label{tab::dataset}
    \begin{tabular}{ccccc}
      \toprule
      dataset & users & items & categories & interactions \\
      \midrule
      Taobao & 82633 & 136710 & 3108 & 4230631 \\
      Beibei & 19140 & 17196 & 110 & 265110 \\
      MSD & 65269 & 40109 & 15 & 2423262 \\
      \bottomrule
    \end{tabular}
\end{table}

For each dataset, we first rank the records according to timestamps, then we select the early 60\% as training set. We divide the last 40\% into two halves. The first 20\% used for validation and hyper-parameter search, and we reserve the last 20\% for performance comparison. To measure the top-K recommendation performance of our proposed method in consideration of both accuracy and diversity, we utilize a bunch of metrics including recall, hit ratio, coverage, entropy and gini index, while the first two metrics are about accuracy and the last three concerns diversity.

\begin{table*}
    \caption{Overall Performance on Taobao dataset and Beibei dataset. (TopK = 300)}
    \label{tab::overall}
    \begin{tabular}{c|ccccc|ccccc}
      \toprule
      dataset & \multicolumn{5}{c|}{Taobao} & \multicolumn{5}{c}{Beibei} \\
      \hline
      metrics & recall & hit ratio & coverage & entropy & gini index & recall & hit ratio & coverage & entropy & gini index \\
      \midrule
      MMR & 0.0544 & 0.0453 & 74.5460 & 3.4931 & 0.5825 & 0.1097 & 0.1036 & 77.016 & 4.0184 & 0.4373 \\
      DUM & 0.0495 & 0.0497 & 126.6621 & 4.1051 & 0.4587 & 0.0746 & 0.0724 & 84.3044 & 4.0389 & 0.4599 \\
      PMF + $\alpha$ + $\beta$ & 0.0473 & 0.0435 & 125.5600 & 4.3725 & 0.4648 & 0.1092 & 0.1054 & 73.4675 & 3.7528 & 0.5127 \\
      DPP & 0.0633 & 0.0485 & 79.1154 & 3.3904 & 0.6096 & 0.0751 & 0.0745 & 69.3416 & 3.7545 & 0.5078 \\
      DGCN & 0.0776 & 0.0783 & 84.6685 & 3.5779 & 0.5583 & 0.1212 & 0.1278 & 71.8546 & 3.7149 & 0.5279 \\
      \bottomrule
    \end{tabular}
\end{table*}

\subsubsection{Baselines} To verify the effectiveness of our proposed DGCN, we compare the performance with several diversification methods as follows:

\begin{itemize}[leftmargin=*]
    \item{\textbf{MMR} \cite{mmr}}: Maximal Marginal Relevance (MMR) is a pioneer work for diversification in search engines and is further adapted to recommendation systems \cite{topic}. This method re-ranks the contents based on greedy algorithms to minimize redundancy.
    \item{\textbf{DUM} \cite{dum}}: This method is also a greedy approach for diversification which aims at maximizing the utility of the items subject to the increase in their diversity.
    \item{\textbf{PMF+$\alpha$+$\beta$} \cite{pmf_alpha_beta}}: This work formalizes the problem as a combination of three aspects: the relevance of the items, the coverage of the user's interest, and the diversity between them. Two hyper-parameters ($\alpha$ and $\beta$) are introduced to balance the three parts.
    \item{\textbf{DPP} \cite{chen2018fast}}: Sourced from mathematics and quantum physics, Determinantal Point Process (DPP) is recently leveraged in machine learning research, serving as an parametric model to provide a diverse subset of items from a larger pool of retrieved items. Several methods \cite{gillenwater2014expectation,chen2018fast,gartrell2017low,gillenwater2019tree,warlop2019tensorized} were proposed to accelerate the computation of DPP.
\end{itemize}

\subsubsection{Parameter Settings} We adopt log loss \cite{ncf} for all methods and fix the embedding size as 32. The AMSGrad \cite{reddi2019convergence} variant of Adam \cite{kingma2014adam} is utilized for optimization. The negative sample rate is set to 4. We train each model until convergence and utilize the early stopping technique to avoid overfitting. We perform grid search to find the best hyper-parameters. Results are averaged over all the users. We implement our proposed method with PyTorch\footnote{https://pytorch.org}, and codes are available at \url{https://github.com/tsinghua-fib-lab/DGCN}.

\subsubsection{Evaluations} Since our work directly uses inner product to estimate the interaction probability, maximum inner product search can be easily integrated into the system, and we use Faiss \cite{faiss} to generate candidates for evaluation which greatly reduces the time cost. During evaluation, we construct a search index (IndexFlatIP\footnote{https://github.com/facebookresearch/faiss} for efficient nearest neighbor search based on inner product) in Faiss using the learned item embeddings, and feed the user embeddings to the search index as query vectors. Items of the maximum inner products with the query vector will be retrieved, and recommendation metrics are further calculated based on the retrieved items. Moreover,evaluations are conducted in batch style and on GPUs for further acceleration. With the help of efficient nearest neighbor search, we successfully reduce the time cost for evaluation to a few seconds.

\subsection{Overall Performance (RQ1)}

We compare our proposed method with several baseline algorithms introduced previously. For each baseline method, we tuned it to be on par with our proposed method in one aspect (accuracy or diversity), and compared the effects of the other aspect, on account of the aforementioned accuracy-diversity tradeoff. Since we incorporate diversification into the matching stage, we use relatively high values of topK (300) to measure the performance of matching. Results on the Taobao and Beibei datasets are illustrated in Table \ref{tab::overall}.
From the results, we have several observations:

\begin{itemize}[leftmargin=*]
	\item \textbf{The accuracy-diversity tradeoff exists widely.} In our experiments, three of the baseline methods (MMR, DUM, PMF+$\alpha$+$\beta$) are based on greedy algorithms, and the other one (DPP) is based on a probability model. Comparing across different methods, generally more diverse methods provide less relevant items. For example, PMF+$\alpha$+$\beta$ achieves much more diverse results than DPP on Taobao dataset with over 50\% relative improvement in terms of coverage, but the accuracy of PMF+$\alpha$+$\beta$ is much inferior to DPP. Similarly, DUM achieves the most diverse results on both datasets, however, the relevance of the recommended items by DUM is greatly damaged by diversification.
	\item \textbf{It is more difficult to balance the two aspects for greedy algorithms.} Although there exist certain hyper-parameters in greedy algorithms to balance the weight for accuracy and diversity, the \textit{slope} or exchange rate of the two aspects tends to be rather large. In other words, greedy algorithms turn out to be more aggressive on diversification which makes the accuracy unacceptable. DUM is such an example which usually generates highly diverse results with relatively poor relevance.
	\item \textbf{Our proposed DGCN achieves a better overall performance.} Generally, DGCN generates more diverse items with reasonable relevance. Compared with DPP, our method attains a better performance with respect to both diversity and accuracy on two datasets. In comparison with MMR, our method outperforms on both diversity and accuracy on Taobao dataset, and performs roughly the same with respect to diversity on Beibei dataset, but with better relevance. Though DUM provides extremely diverse items on two datasets, the relevance of the recommended contents is not qualified enough compared with our method. As for PMF+$\alpha$+$\beta$, our method attains much better accuracy on both datasets, and achieves comparable diversity on Beibei dataset.
\end{itemize}

\begin{figure}[t]
    \centering
    \includegraphics[width=\linewidth]{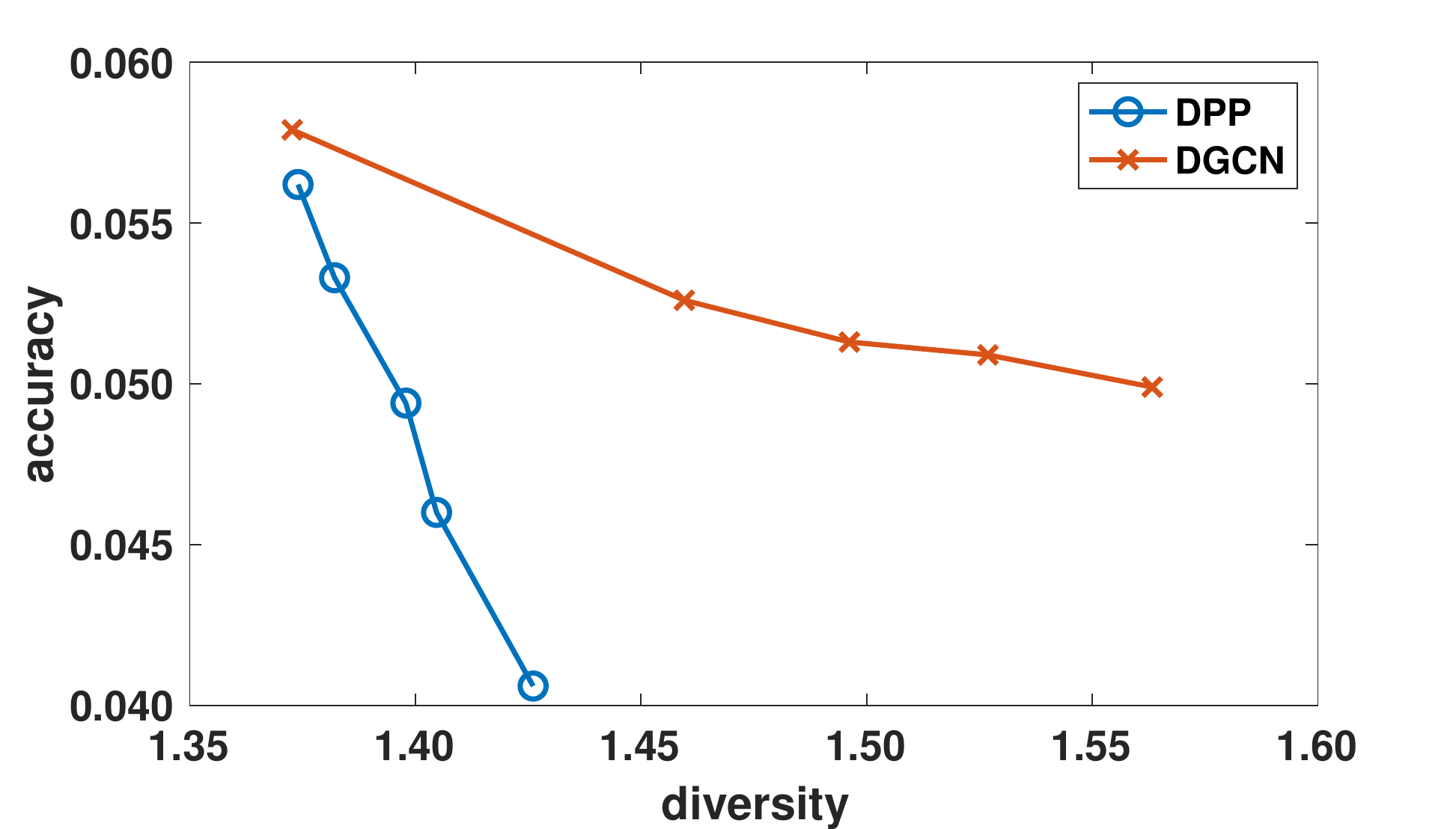}
    \caption{Accuracy-diversity curve of DPP and DGCN on MSD dataset.}
    \vspace{-10px}
    \label{fig::ad-curve}
\end{figure}

To illustrate that our proposed DGCN attains a better overall performance considering both accuracy and diversity, we further conduct experiments on the benchmark MSD dataset, and plot the whole accuracy-diversity curve against the state-of-the-art DPP approach \cite{chen2018fast}.
We tune the tradeoff parameters of DPP and DGCN to obtain recommended items with different accuracy and diversity.
Figure \ref{fig::ad-curve} demonstrates the results on the MSD dataset.
We can observe that the accuracy-diversity curve of the proposed DGCN is closer to the top-right corner than DPP.
In other words, conditioned on equal accuracy, DGCN achieves better diversity than DPP.
Meanwhile, with comparable diversity, the proposed DGCN can provide much more accurate recommendation.
Therefore, the proposed DGCN achieves better overall performance compared with DPP.

\subsection{Study on DGCN (RQ2)}

In this section, we conduct ablation studies on each of our proposed components in DGCN. We compare the performance of our proposed method with and without the special design on rebalanced neighbor discovering, category-boosted negative sampling and adversarial learning. Table \ref{tab::ablation} illustrates the results of GCN without diversificaiton, GCN with only one diversification component and GCN with all the three components (DGCN). We also compare the results with DPP, which is widely adopted for diversified recommendation.

From the results we can observe that each component alone contributes to improve diversity, and combining the three special designs achieves the most diverse recommendation. Specifically, a single GCN without diversification significantly outperforms DPP with respect to accuracy, however, it sacrifices the diversity which can lead to suboptimal user satisfaction. After we incorporate rebalanced neighbor discovering or category-boosted negative sampling, the diversity of our model gets promoted effectively and the model still maintains a relatively high accuracy. Moreover, combining adversarial learning with GCN achieves comparable diversity with DPP, and the recommended items are much more relevant to users' interests. According to the results in Table \ref{tab::ablation}, our proposed DGCN provides the most diverse contents. In summary, without losing the superior capability of GCN, our proposed DGCN is featured of three special designs for diversification on top of GCN, which greatly improves the diversity and also guarantees the relevance of the recommended items.

We employ adversarial learning to distill users' category preferences from item preferences, aiming to make the learned representations to some extent category-free, which in turn increases the probability of recommending items from more diverse categories. We add an adversarial task of item category classification to fulfill this job. The object of the classifier is to maximize the accuracy of predicting items' categories according the learned item embeddings, while the recommendation model aims to fool the classifier as much as possible. In our experiments, we accomplish the task of adversarial training by inserting a Gradient Reversal Layer (GRL). We compare the performance of our model with and without inserting the GRL. Results are shown in Table \ref{tab::adversarial_training}. Without adversarial learning, the learned item embeddings form clusters where items of the same category are near in the embedding space, which is verified by the high category classification accuracy. Most importantly, the diversity of recommendation is rather poor, leading to the problem of information redundancy. After inserting the GRL to perform adversarial learning on category classification, we distill the category information in the embedding space. Thus it becomes much difficult to predict the category from item embeddings. According to the results, the accuracy of category classification drops drastically from 25\% to 6\%, which verifies the effect of the distillation process. Most importantly, with the help of adversarial learning, the recommendation diversity improves significantly with a rather acceptable accuracy.

\begin{table}
    \caption{Ablation study on Taobao dataset.}
    \label{tab::ablation}
    \begin{tabular}{c|cc}
      \toprule
      method & recall & coverage  \\
      \midrule
      DPP & 0.0633 & 79.1154 \\
      \hline
      GCN & 0.1013 & 61.9111 \\
      Rebalance Neighbor Sampling & 0.0939 & 71.2528 \\
      Boost Negative Sampling & 0.0954 & 76.7391 \\
      Adversarial Learning & 0.0846 & 79.0722 \\
      DGCN & 0.0776 & 84.6685  \\
      \bottomrule
    \end{tabular}
\end{table}
\begin{table}
    \caption{Ablation study of adversarial learning on Taobao dataset.}
    \label{tab::adversarial_training}
    \begin{tabular}{c|ccc}
      \toprule
      method & RS acc & RS diversity & Classifier acc \\
      \midrule
      GCN w/o GRL & 0.1041 & 55.6591 & 25\%\\
      GCN w/ GRL & 0.0739 & 80.4894 & 6\% \\
      \bottomrule
    \end{tabular}
\end{table}

\subsection{Trade-off between Accuracy and Diversity (RQ3)}

In the proposed framework, we introduce two hyper-parameters, $\alpha$ and $\beta$, to control the strength of rebalanced neighbor discovering and category-boosted negative sampling. We now investigate whether these two hyper-parameters can be used to perform trade-off between accuracy and diversity.

\subsubsection{\textbf{Rebalanced Neighbor Discovering}}

\begin{figure}[t]
    \centering
    \includegraphics[width=\linewidth]{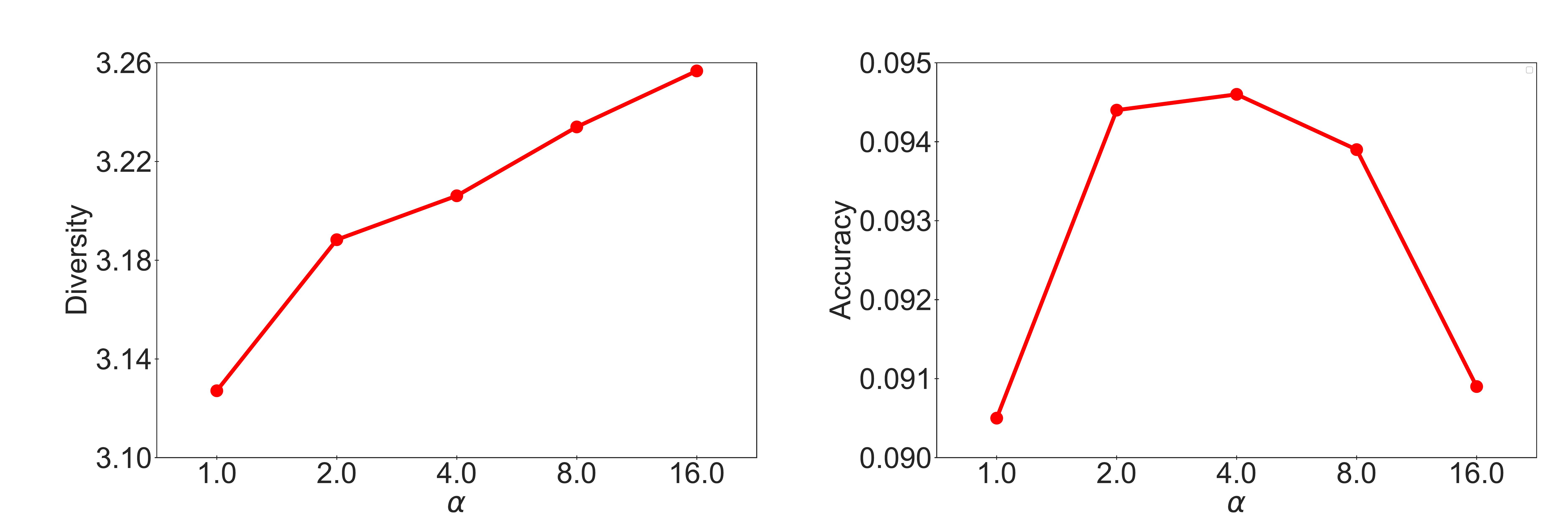}
    \vspace{-10px}
    \caption{Performance on different $\alpha$.}
    \vspace{-10px}
    \label{fig::neighbor}
\end{figure}

We conduct experiments to study the effect of rebalanced neighbor discovering. Results of different values of $\alpha$ are illustrated in Figure \ref{fig::neighbor}. In neighbor discovering, we boost the probability of sampling from those disadvantaged categories, while limit the chances of those dominant categories. With larger $\alpha$, we impose stronger boost on disadvantaged categories and perform more forceful rebalance across categories. As shown in Figure \ref{fig::neighbor}, the diversity of recommendation increases constantly with the growth of $\alpha$. Moreover, the accuracy also gets promoted at relatively lower $\alpha$ and finally drops when $\alpha$ becomes to large, which contradicts the previously introduced accuracy-diversity tradeoff. This phenomenon of attaining improvements in both accuracy and diversity has also been observed in related diversification literatures \cite{wilhelm2018practical,chen2018fast}, which validates that diversification serves as an effective strategy to enhance user satisfaction.

\subsubsection{\textbf{Category-Boosted Negative Sampling}}

\begin{figure}[t]
    \centering
    \includegraphics[width=\linewidth]{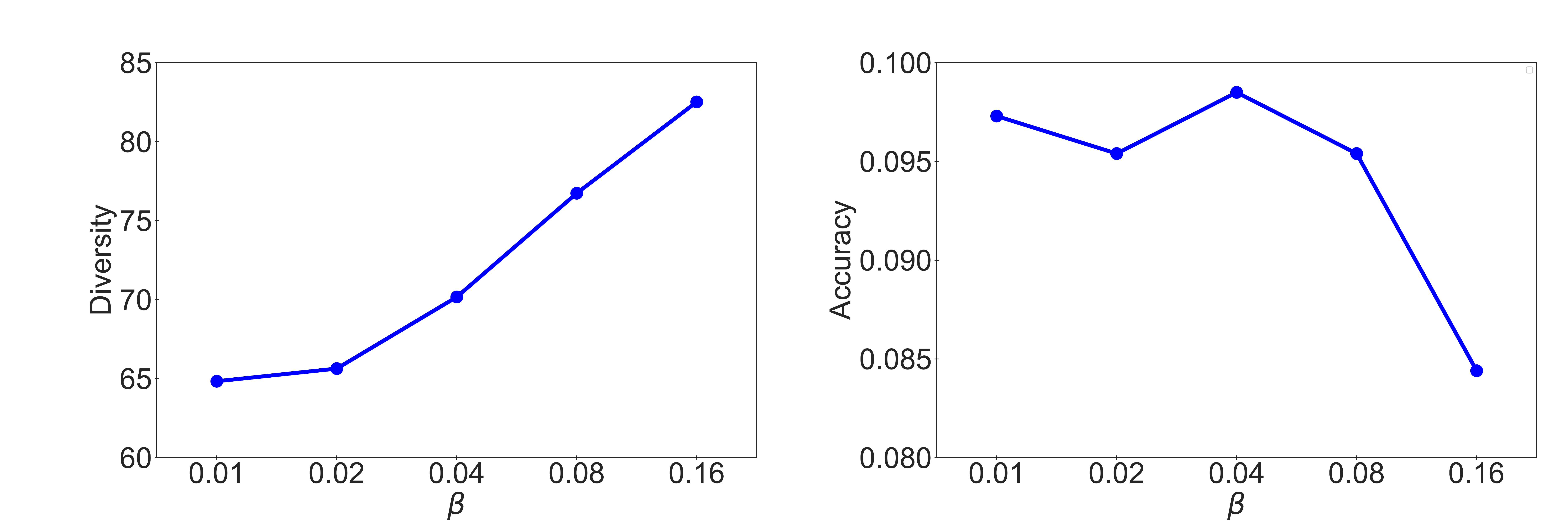}
    \vspace{-10px}
    \caption{Performance on different $\beta$.}
    \vspace{-10px}
    \label{fig::negative}
\end{figure}

Experiments are conducted over different values of $\beta$. In our work, we make adjustments to the negative sampling process, where we aim to find those\textit{similar but negative} items. Specifically, we sample from positive categories with probability $\beta$ which is much larger than the probability by random sampling. By selecting more negative items from positive categories, the learned representations capture users' interest across categories and items of more diverse categories are recommended. Figure \ref{fig::negative} illustrates the performance on different $\beta$ with respect to accuracy and diversity. Similar to the previous experiments on neighbor discovering, the diversity of recommendation improves as we increase the probability of sampling from similar items. In addition, the accuracy is rather stable on small $\beta$ and decreases when it gets too large. Through category-boosted negative sampling, our proposed DGCN provides more diverse items and guarantees the relevance of the recommended contents.


In summary, we conduct extensive experiments to evaluate our proposed DGCN with special emphasis on diversification. Overall performance on real-world datasets confirms the effectiveness of our method on improving diversity. Ablation studies of DGCN verify the function of each component. Further experiments demonstrate that trade-off between accuracy and diversity can be smoothly performed by tuning the introduced hyper-parameters.

\section{Related Work} \label{sec::related}

\noindent\textbf{Diversified Recommendation}
Research on diversification in recommendation was first introduced by Ziegler \textit{et al.} \cite{topic}, which leveraged a greedy algorithm \cite{mmr} from the field of information retrieval. After that, a series of post processing methods were proposed to diversify the recommendation results. Qin \textit{et al.} \cite{entropy} tackled the problem by performing a linear combination of the rating function and a entropy regularizer. Ashkan \text{et al.} \cite{dum} replaced the weighted sum object in greedy solutions with multiplication, thus removed a tuned parameter for balancing utility and diversity. Sha \text{et al.} \cite{pmf_alpha_beta} developed a framework combining relevance, coverage of user interests and diversity. Two hyper-parameters, $\alpha$ and $\beta$, were introduced to balance the object. Other than reranking based methods, a series of solutions called learning to rank (LTR) were proposed to generate the recommended list directly. Cheng \textit{et al.} \cite{dcf} developed a learning-based diversification method by coupling the recommendation model with a structural SVM \cite{tsochantaridis2005large}. Li \textit{et al.}\cite{li2017learning} proposed a ranking model and utilized a score function with the form of the product of the estimated interaction probability and category preference. Factorized category features were leveraged for optimization. Recently, Determinantal Point Process (DPP) was introduced to recommendation to generate diverse items. Several algorithms \cite{gillenwater2014expectation,chen2018fast,gartrell2017low,gillenwater2019tree,warlop2019tensorized,wilhelm2018practical} were proposed to reduce the heavy computation of DPP with the help of EM algorithms, greedy algorithms or tensor factorization.
Unlike existing works that mainly perform diversification after matching, our proposed method combines diversification and matching with an end-to-end GCN model.

\noindent\textbf{GCN based Recommendation}
In recent years, Graph Convolutional Networks (GCN) has made great progress in network representation learning tasks including node classification and link prediction \cite{gcn,graphsage,powerful}. Several works \cite{gcmc,pinsage,NGCF,KGAT,wang2019knowledge,zheng2020price} have taken advantage of GCN to learn more robust latent representations for users and items in recommendation systems. With more advanced capacity in learning graph representations, GCN has also been shown effective and efficient to be deployed in web-scale recommendation applications \cite{pinsage}. By applying GCN \cite{gcn} on the user-item interaction bipartite graph, Berg \textit{et al.} \cite{gcmc} transformed the matrix completion task in recommendation to link prediction on the graph. Ying \textit{et al.} \cite{pinsage} further extended the inductive learning idea introduced in \cite{graphsage} to practical recommendation scenario, and proposed a scalable algorithm called PinSage. Both offline and online evaluations confirm the effectiveness of GCN in modeling user preference. Wang \textit{et al.} \cite{NGCF} developed a framework which performs embedding propagation on the user-item integration graph based on GCN to model the high-order connectivity. 
Experimental results illustrated that the accuracy of recommendation has been successfully improved by utilizing GCN to perform representation learning. However, how diversity is impacted by the complicated GCN model remains uncertain. In our work, we focus on diversified recommendation with the help of GCN.
\section{Conclusion} \label{sec::con}

In this work, we investigated existing diversification solutions and pointed out the challenge that the decoupled design of diversification and matching could lead to suboptimal performance. Based on our analysis, we aimed to push the diversification process upwards into the matching stage, and proposed an end-to-end diversified recommendation model based on GCN with several special designs on diversity. We conducted extensive experiments on real-world datasets. Experimental results validated the effectiveness of our proposed method on improving diversity. Further ablation studies confirmed that our proposed DGCN provides diverse and relevant contents to meet users' needs.

Although the accuracy-diversity tradeoff still exists in our proposed method, it has been shown that improving the diversity does not necessarily lead to inferior accuracy \cite{szpektor2013relevance,wilhelm2018practical,chen2018fast}. Especially in online scenarios, promoting the diversity of the recommended contents yields substantial increases in user engagement \cite{wilhelm2018practical}. The conflict of diversity and accuracy to some extent results from the differences between offline evaluation and online evaluation, as well as the causality of the system \cite{sculley2015hidden}, which we believe are quite interesting and important research questions. One step further, distinct users might regard diversity differently, and the diversification process might also be personalized (\textit{i.e.} personalized personalization) \cite{amatriain2016past}, which is also a promising future direction.

\begin{acks}
This work was supported in part by the National Natural Science Foundation of China under 61941117, U1936217,  61971267, 61972223, 61861136003.
\end{acks}

\bibliographystyle{ACM-Reference-Format}
\bibliography{bibliography}


\begin{thebibliography}{69}


\ifx \showCODEN    \undefined \def \showCODEN     #1{\unskip}     \fi
\ifx \showDOI      \undefined \def \showDOI       #1{#1}\fi
\ifx \showISBNx    \undefined \def \showISBNx     #1{\unskip}     \fi
\ifx \showISBNxiii \undefined \def \showISBNxiii  #1{\unskip}     \fi
\ifx \showISSN     \undefined \def \showISSN      #1{\unskip}     \fi
\ifx \showLCCN     \undefined \def \showLCCN      #1{\unskip}     \fi
\ifx \shownote     \undefined \def \shownote      #1{#1}          \fi
\ifx \showarticletitle \undefined \def \showarticletitle #1{#1}   \fi
\ifx \showURL      \undefined \def \showURL       {\relax}        \fi
\providecommand\bibfield[2]{#2}
\providecommand\bibinfo[2]{#2}
\providecommand\natexlab[1]{#1}
\providecommand\showeprint[2][]{arXiv:#2}

\bibitem[\protect\citeauthoryear{Amatriain and Basilico}{Amatriain and
  Basilico}{2016}]%
        {amatriain2016past}
\bibfield{author}{\bibinfo{person}{Xavier Amatriain} {and}
  \bibinfo{person}{Justin Basilico}.} \bibinfo{year}{2016}\natexlab{}.
\newblock \showarticletitle{Past, present, and future of recommender systems:
  An industry perspective}. In \bibinfo{booktitle}{\emph{Proceedings of the
  10th ACM conference on recommender systems}}. \bibinfo{pages}{211--214}.
\newblock


\bibitem[\protect\citeauthoryear{Antikacioglu and Ravi}{Antikacioglu and
  Ravi}{2017}]%
        {antikacioglu2017post}
\bibfield{author}{\bibinfo{person}{Arda Antikacioglu} {and} \bibinfo{person}{R
  Ravi}.} \bibinfo{year}{2017}\natexlab{}.
\newblock \showarticletitle{Post processing recommender systems for diversity}.
  In \bibinfo{booktitle}{\emph{Proceedings of the 23rd ACM SIGKDD International
  Conference on Knowledge Discovery and Data Mining}}. ACM,
  \bibinfo{pages}{707--716}.
\newblock


\bibitem[\protect\citeauthoryear{Ashkan, Kveton, Berkovsky, and Wen}{Ashkan
  et~al\mbox{.}}{2015}]%
        {dum}
\bibfield{author}{\bibinfo{person}{Azin Ashkan}, \bibinfo{person}{Branislav
  Kveton}, \bibinfo{person}{Shlomo Berkovsky}, {and} \bibinfo{person}{Zheng
  Wen}.} \bibinfo{year}{2015}\natexlab{}.
\newblock \showarticletitle{Optimal greedy diversity for recommendation}. In
  \bibinfo{booktitle}{\emph{Twenty-Fourth International Joint Conference on
  Artificial Intelligence}}.
\newblock


\bibitem[\protect\citeauthoryear{Berg, Kipf, and Welling}{Berg
  et~al\mbox{.}}{2017}]%
        {gcmc}
\bibfield{author}{\bibinfo{person}{Rianne van~den Berg},
  \bibinfo{person}{Thomas~N Kipf}, {and} \bibinfo{person}{Max Welling}.}
  \bibinfo{year}{2017}\natexlab{}.
\newblock \showarticletitle{Graph convolutional matrix completion}.
\newblock \bibinfo{journal}{\emph{arXiv preprint arXiv:1706.02263}}
  (\bibinfo{year}{2017}).
\newblock


\bibitem[\protect\citeauthoryear{Bertin-Mahieux, Ellis, Whitman, and
  Lamere}{Bertin-Mahieux et~al\mbox{.}}{2011}]%
        {Bertin-Mahieux2011}
\bibfield{author}{\bibinfo{person}{Thierry Bertin-Mahieux},
  \bibinfo{person}{Daniel~P.W. Ellis}, \bibinfo{person}{Brian Whitman}, {and}
  \bibinfo{person}{Paul Lamere}.} \bibinfo{year}{2011}\natexlab{}.
\newblock \showarticletitle{The Million Song Dataset}. In
  \bibinfo{booktitle}{\emph{{Proceedings of the 12th International Conference
  on Music Information Retrieval ({ISMIR} 2011)}}}.
\newblock


\bibitem[\protect\citeauthoryear{Boim, Milo, and Novgorodov}{Boim
  et~al\mbox{.}}{2011}]%
        {direc}
\bibfield{author}{\bibinfo{person}{Rubi Boim}, \bibinfo{person}{Tova Milo},
  {and} \bibinfo{person}{Slava Novgorodov}.} \bibinfo{year}{2011}\natexlab{}.
\newblock \showarticletitle{Diversification and refinement in collaborative
  filtering recommender}. In \bibinfo{booktitle}{\emph{Proceedings of the 20th
  ACM international conference on Information and knowledge management}}. ACM,
  \bibinfo{pages}{739--744}.
\newblock


\bibitem[\protect\citeauthoryear{Borodin, Lee, and Ye}{Borodin
  et~al\mbox{.}}{2012}]%
        {msd}
\bibfield{author}{\bibinfo{person}{Allan Borodin}, \bibinfo{person}{Hyun~Chul
  Lee}, {and} \bibinfo{person}{Yuli Ye}.} \bibinfo{year}{2012}\natexlab{}.
\newblock \showarticletitle{Max-sum diversification, monotone submodular
  functions and dynamic updates}. In \bibinfo{booktitle}{\emph{Proceedings of
  the 31st ACM SIGMOD-SIGACT-SIGAI symposium on Principles of Database
  Systems}}. ACM, \bibinfo{pages}{155--166}.
\newblock


\bibitem[\protect\citeauthoryear{Carbonell and Goldstein}{Carbonell and
  Goldstein}{1998}]%
        {mmr}
\bibfield{author}{\bibinfo{person}{Jaime~G Carbonell} {and}
  \bibinfo{person}{Jade Goldstein}.} \bibinfo{year}{1998}\natexlab{}.
\newblock \showarticletitle{The use of MMR, diversity-based reranking for
  reordering documents and producing summaries.}. In
  \bibinfo{booktitle}{\emph{SIGIR}}, Vol.~\bibinfo{volume}{98}.
  \bibinfo{pages}{335--336}.
\newblock


\bibitem[\protect\citeauthoryear{Chen, Zhang, He, Nie, Liu, and Chua}{Chen
  et~al\mbox{.}}{2017}]%
        {chen2017attentive}
\bibfield{author}{\bibinfo{person}{Jingyuan Chen}, \bibinfo{person}{Hanwang
  Zhang}, \bibinfo{person}{Xiangnan He}, \bibinfo{person}{Liqiang Nie},
  \bibinfo{person}{Wei Liu}, {and} \bibinfo{person}{Tat-Seng Chua}.}
  \bibinfo{year}{2017}\natexlab{}.
\newblock \showarticletitle{Attentive collaborative filtering: Multimedia
  recommendation with item-and component-level attention}. In
  \bibinfo{booktitle}{\emph{Proceedings of the 40th International ACM SIGIR
  conference on Research and Development in Information Retrieval}}. ACM,
  \bibinfo{pages}{335--344}.
\newblock


\bibitem[\protect\citeauthoryear{Chen, Wu, and He}{Chen et~al\mbox{.}}{2016}]%
        {chen2016personality}
\bibfield{author}{\bibinfo{person}{Li Chen}, \bibinfo{person}{Wen Wu}, {and}
  \bibinfo{person}{Liang He}.} \bibinfo{year}{2016}\natexlab{}.
\newblock \showarticletitle{Personality and recommendation diversity}.
\newblock In \bibinfo{booktitle}{\emph{Emotions and Personality in Personalized
  Services}}. \bibinfo{publisher}{Springer}, \bibinfo{pages}{201--225}.
\newblock


\bibitem[\protect\citeauthoryear{Chen, Zhang, and Zhou}{Chen
  et~al\mbox{.}}{2018}]%
        {chen2018fast}
\bibfield{author}{\bibinfo{person}{Laming Chen}, \bibinfo{person}{Guoxin
  Zhang}, {and} \bibinfo{person}{Eric Zhou}.} \bibinfo{year}{2018}\natexlab{}.
\newblock \showarticletitle{Fast greedy map inference for determinantal point
  process to improve recommendation diversity}. In
  \bibinfo{booktitle}{\emph{Advances in Neural Information Processing
  Systems}}. \bibinfo{pages}{5622--5633}.
\newblock


\bibitem[\protect\citeauthoryear{Cheng, Wang, Ma, Sun, and Xiong}{Cheng
  et~al\mbox{.}}{2017}]%
        {dcf}
\bibfield{author}{\bibinfo{person}{Peizhe Cheng}, \bibinfo{person}{Shuaiqiang
  Wang}, \bibinfo{person}{Jun Ma}, \bibinfo{person}{Jiankai Sun}, {and}
  \bibinfo{person}{Hui Xiong}.} \bibinfo{year}{2017}\natexlab{}.
\newblock \showarticletitle{Learning to recommend accurate and diverse items}.
  In \bibinfo{booktitle}{\emph{Proceedings of the 26th international conference
  on World Wide Web}}. International World Wide Web Conferences Steering
  Committee, \bibinfo{pages}{183--192}.
\newblock


\bibitem[\protect\citeauthoryear{Covington, Adams, and Sargin}{Covington
  et~al\mbox{.}}{2016}]%
        {youtube}
\bibfield{author}{\bibinfo{person}{Paul Covington}, \bibinfo{person}{Jay
  Adams}, {and} \bibinfo{person}{Emre Sargin}.}
  \bibinfo{year}{2016}\natexlab{}.
\newblock \showarticletitle{Deep neural networks for youtube recommendations}.
  In \bibinfo{booktitle}{\emph{Proceedings of the 10th ACM conference on
  recommender systems}}. ACM, \bibinfo{pages}{191--198}.
\newblock


\bibitem[\protect\citeauthoryear{Ding, Feng, He, Yu, Li, and Jin}{Ding
  et~al\mbox{.}}{2018}]%
        {ding2018www}
\bibfield{author}{\bibinfo{person}{Jingtao Ding}, \bibinfo{person}{Fuli Feng},
  \bibinfo{person}{Xiangnan He}, \bibinfo{person}{Guanghui Yu},
  \bibinfo{person}{Yong Li}, {and} \bibinfo{person}{Depeng Jin}.}
  \bibinfo{year}{2018}\natexlab{}.
\newblock \showarticletitle{An improved sampler for bayesian personalized
  ranking by leveraging view data}. In \bibinfo{booktitle}{\emph{Companion
  Proceedings of the The Web Conference 2018}}. International World Wide Web
  Conferences Steering Committee, \bibinfo{pages}{13--14}.
\newblock


\bibitem[\protect\citeauthoryear{Ding, Quan, He, Li, and Jin}{Ding
  et~al\mbox{.}}{2019}]%
        {ding2019reinforced}
\bibfield{author}{\bibinfo{person}{Jingtao Ding}, \bibinfo{person}{Yuhan Quan},
  \bibinfo{person}{Xiangnan He}, \bibinfo{person}{Yong Li}, {and}
  \bibinfo{person}{Depeng Jin}.} \bibinfo{year}{2019}\natexlab{}.
\newblock \showarticletitle{Reinforced Negative Sampling for Recommendation
  with Exposure Data}. IJCAI.
\newblock


\bibitem[\protect\citeauthoryear{Eksombatchai, Jindal, Liu, Liu, Sharma,
  Sugnet, Ulrich, and Leskovec}{Eksombatchai et~al\mbox{.}}{2018}]%
        {pixie}
\bibfield{author}{\bibinfo{person}{Chantat Eksombatchai},
  \bibinfo{person}{Pranav Jindal}, \bibinfo{person}{Jerry~Zitao Liu},
  \bibinfo{person}{Yuchen Liu}, \bibinfo{person}{Rahul Sharma},
  \bibinfo{person}{Charles Sugnet}, \bibinfo{person}{Mark Ulrich}, {and}
  \bibinfo{person}{Jure Leskovec}.} \bibinfo{year}{2018}\natexlab{}.
\newblock \showarticletitle{Pixie: A system for recommending 3+ billion items
  to 200+ million users in real-time}. In \bibinfo{booktitle}{\emph{Proceedings
  of the 2018 World Wide Web Conference}}. International World Wide Web
  Conferences Steering Committee, \bibinfo{pages}{1775--1784}.
\newblock


\bibitem[\protect\citeauthoryear{Ganin and Lempitsky}{Ganin and
  Lempitsky}{2014}]%
        {DAN}
\bibfield{author}{\bibinfo{person}{Yaroslav Ganin} {and}
  \bibinfo{person}{Victor Lempitsky}.} \bibinfo{year}{2014}\natexlab{}.
\newblock \showarticletitle{Unsupervised domain adaptation by backpropagation}.
\newblock \bibinfo{journal}{\emph{arXiv preprint arXiv:1409.7495}}
  (\bibinfo{year}{2014}).
\newblock


\bibitem[\protect\citeauthoryear{Gao, He, Gan, Chen, Feng, Li, Chua, and
  Jin}{Gao et~al\mbox{.}}{2018}]%
        {gao2018learning}
\bibfield{author}{\bibinfo{person}{Chen Gao}, \bibinfo{person}{Xiangnan He},
  \bibinfo{person}{Dahua Gan}, \bibinfo{person}{Xiangning Chen},
  \bibinfo{person}{Fuli Feng}, \bibinfo{person}{Yong Li},
  \bibinfo{person}{Tat-Seng Chua}, {and} \bibinfo{person}{Depeng Jin}.}
  \bibinfo{year}{2018}\natexlab{}.
\newblock \showarticletitle{Learning recommender systems from multi-behavior
  data}.
\newblock \bibinfo{journal}{\emph{arXiv preprint arXiv:1809.08161}}
  (\bibinfo{year}{2018}).
\newblock


\bibitem[\protect\citeauthoryear{Gartrell, Paquet, and Koenigstein}{Gartrell
  et~al\mbox{.}}{2017}]%
        {gartrell2017low}
\bibfield{author}{\bibinfo{person}{Mike Gartrell}, \bibinfo{person}{Ulrich
  Paquet}, {and} \bibinfo{person}{Noam Koenigstein}.}
  \bibinfo{year}{2017}\natexlab{}.
\newblock \showarticletitle{Low-rank factorization of determinantal point
  processes}. In \bibinfo{booktitle}{\emph{Thirty-First AAAI Conference on
  Artificial Intelligence}}.
\newblock


\bibitem[\protect\citeauthoryear{Gautier, Polito, Bardenet, and Valko}{Gautier
  et~al\mbox{.}}{2018}]%
        {dppy}
\bibfield{author}{\bibinfo{person}{Guillaume Gautier},
  \bibinfo{person}{Guillermo Polito}, \bibinfo{person}{R{\'{e}}mi Bardenet},
  {and} \bibinfo{person}{Michal Valko}.} \bibinfo{year}{2018}\natexlab{}.
\newblock \showarticletitle{{DPPy: Sampling Determinantal Point Processes with
  Python}}.
\newblock \bibinfo{journal}{\emph{ArXiv e-prints}} (\bibinfo{year}{2018}).
\newblock
\showeprint[arxiv]{1809.07258}
\urldef\tempurl%
\url{http://arxiv.org/abs/1809.07258}
\showURL{%
\tempurl}
\newblock
\shownote{Code at http://github.com/guilgautier/DPPy/ Documentation at
  http://dppy.readthedocs.io/.}


\bibitem[\protect\citeauthoryear{Gillenwater, Kulesza, Mariet, and
  Vassilvtiskii}{Gillenwater et~al\mbox{.}}{2019}]%
        {gillenwater2019tree}
\bibfield{author}{\bibinfo{person}{Jennifer Gillenwater}, \bibinfo{person}{Alex
  Kulesza}, \bibinfo{person}{Zelda Mariet}, {and} \bibinfo{person}{Sergei
  Vassilvtiskii}.} \bibinfo{year}{2019}\natexlab{}.
\newblock \showarticletitle{A Tree-Based Method for Fast Repeated Sampling of
  Determinantal Point Processes}. In \bibinfo{booktitle}{\emph{International
  Conference on Machine Learning}}. \bibinfo{pages}{2260--2268}.
\newblock


\bibitem[\protect\citeauthoryear{Gillenwater, Kulesza, Fox, and
  Taskar}{Gillenwater et~al\mbox{.}}{2014}]%
        {gillenwater2014expectation}
\bibfield{author}{\bibinfo{person}{Jennifer~A Gillenwater},
  \bibinfo{person}{Alex Kulesza}, \bibinfo{person}{Emily Fox}, {and}
  \bibinfo{person}{Ben Taskar}.} \bibinfo{year}{2014}\natexlab{}.
\newblock \showarticletitle{Expectation-maximization for learning determinantal
  point processes}. In \bibinfo{booktitle}{\emph{Advances in Neural Information
  Processing Systems}}. \bibinfo{pages}{3149--3157}.
\newblock


\bibitem[\protect\citeauthoryear{Goodfellow, Pouget-Abadie, Mirza, Xu,
  Warde-Farley, Ozair, Courville, and Bengio}{Goodfellow et~al\mbox{.}}{2014}]%
        {gan}
\bibfield{author}{\bibinfo{person}{Ian Goodfellow}, \bibinfo{person}{Jean
  Pouget-Abadie}, \bibinfo{person}{Mehdi Mirza}, \bibinfo{person}{Bing Xu},
  \bibinfo{person}{David Warde-Farley}, \bibinfo{person}{Sherjil Ozair},
  \bibinfo{person}{Aaron Courville}, {and} \bibinfo{person}{Yoshua Bengio}.}
  \bibinfo{year}{2014}\natexlab{}.
\newblock \showarticletitle{Generative adversarial nets}. In
  \bibinfo{booktitle}{\emph{Advances in neural information processing
  systems}}. \bibinfo{pages}{2672--2680}.
\newblock


\bibitem[\protect\citeauthoryear{Grbovic and Cheng}{Grbovic and Cheng}{2018}]%
        {airbnb}
\bibfield{author}{\bibinfo{person}{Mihajlo Grbovic} {and}
  \bibinfo{person}{Haibin Cheng}.} \bibinfo{year}{2018}\natexlab{}.
\newblock \showarticletitle{Real-time personalization using embeddings for
  search ranking at airbnb}. In \bibinfo{booktitle}{\emph{Proceedings of the
  24th ACM SIGKDD International Conference on Knowledge Discovery \& Data
  Mining}}. ACM, \bibinfo{pages}{311--320}.
\newblock


\bibitem[\protect\citeauthoryear{Guo, Tang, Ye, Li, and He}{Guo
  et~al\mbox{.}}{2017}]%
        {guo2017deepfm}
\bibfield{author}{\bibinfo{person}{Huifeng Guo}, \bibinfo{person}{Ruiming
  Tang}, \bibinfo{person}{Yunming Ye}, \bibinfo{person}{Zhenguo Li}, {and}
  \bibinfo{person}{Xiuqiang He}.} \bibinfo{year}{2017}\natexlab{}.
\newblock \showarticletitle{DeepFM: a factorization-machine based neural
  network for CTR prediction}.
\newblock \bibinfo{journal}{\emph{arXiv preprint arXiv:1703.04247}}
  (\bibinfo{year}{2017}).
\newblock


\bibitem[\protect\citeauthoryear{Hamilton, Ying, and Leskovec}{Hamilton
  et~al\mbox{.}}{2017}]%
        {graphsage}
\bibfield{author}{\bibinfo{person}{Will Hamilton}, \bibinfo{person}{Zhitao
  Ying}, {and} \bibinfo{person}{Jure Leskovec}.}
  \bibinfo{year}{2017}\natexlab{}.
\newblock \showarticletitle{Inductive representation learning on large graphs}.
  In \bibinfo{booktitle}{\emph{Advances in Neural Information Processing
  Systems}}. \bibinfo{pages}{1024--1034}.
\newblock


\bibitem[\protect\citeauthoryear{He and McAuley}{He and McAuley}{2016}]%
        {he2016vbpr}
\bibfield{author}{\bibinfo{person}{Ruining He} {and} \bibinfo{person}{Julian
  McAuley}.} \bibinfo{year}{2016}\natexlab{}.
\newblock \showarticletitle{VBPR: visual bayesian personalized ranking from
  implicit feedback}. In \bibinfo{booktitle}{\emph{Thirtieth AAAI Conference on
  Artificial Intelligence}}.
\newblock


\bibitem[\protect\citeauthoryear{He and Chua}{He and Chua}{2017}]%
        {he2017neural}
\bibfield{author}{\bibinfo{person}{Xiangnan He} {and} \bibinfo{person}{Tat-Seng
  Chua}.} \bibinfo{year}{2017}\natexlab{}.
\newblock \showarticletitle{Neural factorization machines for sparse predictive
  analytics}. In \bibinfo{booktitle}{\emph{Proceedings of the 40th
  International ACM SIGIR conference on Research and Development in Information
  Retrieval}}. ACM, \bibinfo{pages}{355--364}.
\newblock


\bibitem[\protect\citeauthoryear{He, Liao, Zhang, Nie, Hu, and Chua}{He
  et~al\mbox{.}}{2017}]%
        {ncf}
\bibfield{author}{\bibinfo{person}{Xiangnan He}, \bibinfo{person}{Lizi Liao},
  \bibinfo{person}{Hanwang Zhang}, \bibinfo{person}{Liqiang Nie},
  \bibinfo{person}{Xia Hu}, {and} \bibinfo{person}{Tat-Seng Chua}.}
  \bibinfo{year}{2017}\natexlab{}.
\newblock \showarticletitle{Neural collaborative filtering}. In
  \bibinfo{booktitle}{\emph{Proceedings of the 26th international conference on
  world wide web}}. International World Wide Web Conferences Steering
  Committee, \bibinfo{pages}{173--182}.
\newblock


\bibitem[\protect\citeauthoryear{Jiang, Liu, Fu, Wu, and Zhang}{Jiang
  et~al\mbox{.}}{2018}]%
        {jiang2018recommendation}
\bibfield{author}{\bibinfo{person}{Zhengshen Jiang}, \bibinfo{person}{Hongzhi
  Liu}, \bibinfo{person}{Bin Fu}, \bibinfo{person}{Zhonghai Wu}, {and}
  \bibinfo{person}{Tao Zhang}.} \bibinfo{year}{2018}\natexlab{}.
\newblock \showarticletitle{Recommendation in heterogeneous information
  networks based on generalized random walk model and bayesian personalized
  ranking}. In \bibinfo{booktitle}{\emph{Proceedings of the Eleventh ACM
  International Conference on Web Search and Data Mining}}. ACM,
  \bibinfo{pages}{288--296}.
\newblock


\bibitem[\protect\citeauthoryear{Johnson, Douze, and J{\'e}gou}{Johnson
  et~al\mbox{.}}{2017}]%
        {faiss}
\bibfield{author}{\bibinfo{person}{Jeff Johnson}, \bibinfo{person}{Matthijs
  Douze}, {and} \bibinfo{person}{Herv{\'e} J{\'e}gou}.}
  \bibinfo{year}{2017}\natexlab{}.
\newblock \showarticletitle{Billion-scale similarity search with GPUs}.
\newblock \bibinfo{journal}{\emph{arXiv preprint arXiv:1702.08734}}
  (\bibinfo{year}{2017}).
\newblock


\bibitem[\protect\citeauthoryear{Kapoor, Kumar, Terveen, Konstan, and
  Schrater}{Kapoor et~al\mbox{.}}{2015}]%
        {kapoor2015like}
\bibfield{author}{\bibinfo{person}{Komal Kapoor}, \bibinfo{person}{Vikas
  Kumar}, \bibinfo{person}{Loren Terveen}, \bibinfo{person}{Joseph~A Konstan},
  {and} \bibinfo{person}{Paul Schrater}.} \bibinfo{year}{2015}\natexlab{}.
\newblock \showarticletitle{I like to explore sometimes: Adapting to dynamic
  user novelty preferences}. In \bibinfo{booktitle}{\emph{Proceedings of the
  9th ACM Conference on Recommender Systems}}. ACM, \bibinfo{pages}{19--26}.
\newblock


\bibitem[\protect\citeauthoryear{Kingma and Ba}{Kingma and Ba}{2014}]%
        {kingma2014adam}
\bibfield{author}{\bibinfo{person}{Diederik~P Kingma} {and}
  \bibinfo{person}{Jimmy Ba}.} \bibinfo{year}{2014}\natexlab{}.
\newblock \showarticletitle{Adam: A method for stochastic optimization}.
\newblock \bibinfo{journal}{\emph{arXiv preprint arXiv:1412.6980}}
  (\bibinfo{year}{2014}).
\newblock


\bibitem[\protect\citeauthoryear{Kipf and Welling}{Kipf and Welling}{2016}]%
        {gcn}
\bibfield{author}{\bibinfo{person}{Thomas~N Kipf} {and} \bibinfo{person}{Max
  Welling}.} \bibinfo{year}{2016}\natexlab{}.
\newblock \showarticletitle{Semi-supervised classification with graph
  convolutional networks}.
\newblock \bibinfo{journal}{\emph{arXiv preprint arXiv:1609.02907}}
  (\bibinfo{year}{2016}).
\newblock


\bibitem[\protect\citeauthoryear{Koren}{Koren}{2008}]%
        {MF}
\bibfield{author}{\bibinfo{person}{Yehuda Koren}.}
  \bibinfo{year}{2008}\natexlab{}.
\newblock \showarticletitle{Factorization meets the neighborhood: a
  multifaceted collaborative filtering model}. In
  \bibinfo{booktitle}{\emph{Proceedings of the 14th ACM SIGKDD international
  conference on Knowledge discovery and data mining}}. ACM,
  \bibinfo{pages}{426--434}.
\newblock


\bibitem[\protect\citeauthoryear{Li, Zhou, Zhang, Zhang, and Lan}{Li
  et~al\mbox{.}}{2017}]%
        {li2017learning}
\bibfield{author}{\bibinfo{person}{Shuang Li}, \bibinfo{person}{Yuezhi Zhou},
  \bibinfo{person}{Di Zhang}, \bibinfo{person}{Yaoxue Zhang}, {and}
  \bibinfo{person}{Xiang Lan}.} \bibinfo{year}{2017}\natexlab{}.
\newblock \showarticletitle{Learning to Diversify Recommendations Based on
  Matrix Factorization}. In \bibinfo{booktitle}{\emph{2017 IEEE 15th Intl Conf
  on Dependable, Autonomic and Secure Computing, 15th Intl Conf on Pervasive
  Intelligence and Computing, 3rd Intl Conf on Big Data Intelligence and
  Computing and Cyber Science and Technology Congress
  (DASC/PiCom/DataCom/CyberSciTech)}}. IEEE, \bibinfo{pages}{68--74}.
\newblock


\bibitem[\protect\citeauthoryear{Lian, Zhou, Zhang, Chen, Xie, and Sun}{Lian
  et~al\mbox{.}}{2018}]%
        {lian2018xdeepfm}
\bibfield{author}{\bibinfo{person}{Jianxun Lian}, \bibinfo{person}{Xiaohuan
  Zhou}, \bibinfo{person}{Fuzheng Zhang}, \bibinfo{person}{Zhongxia Chen},
  \bibinfo{person}{Xing Xie}, {and} \bibinfo{person}{Guangzhong Sun}.}
  \bibinfo{year}{2018}\natexlab{}.
\newblock \showarticletitle{xdeepfm: Combining explicit and implicit feature
  interactions for recommender systems}. In
  \bibinfo{booktitle}{\emph{Proceedings of the 24th ACM SIGKDD International
  Conference on Knowledge Discovery \& Data Mining}}. ACM,
  \bibinfo{pages}{1754--1763}.
\newblock


\bibitem[\protect\citeauthoryear{Linden, Smith, and York}{Linden
  et~al\mbox{.}}{2003}]%
        {amazon}
\bibfield{author}{\bibinfo{person}{Greg Linden}, \bibinfo{person}{Brent Smith},
  {and} \bibinfo{person}{Jeremy York}.} \bibinfo{year}{2003}\natexlab{}.
\newblock \showarticletitle{Amazon. com recommendations: Item-to-item
  collaborative filtering}.
\newblock \bibinfo{journal}{\emph{IEEE Internet computing}} \bibinfo{number}{1}
  (\bibinfo{year}{2003}), \bibinfo{pages}{76--80}.
\newblock


\bibitem[\protect\citeauthoryear{Mikolov, Sutskever, Chen, Corrado, and
  Dean}{Mikolov et~al\mbox{.}}{2013}]%
        {word2vec}
\bibfield{author}{\bibinfo{person}{Tomas Mikolov}, \bibinfo{person}{Ilya
  Sutskever}, \bibinfo{person}{Kai Chen}, \bibinfo{person}{Greg~S Corrado},
  {and} \bibinfo{person}{Jeff Dean}.} \bibinfo{year}{2013}\natexlab{}.
\newblock \showarticletitle{Distributed representations of words and phrases
  and their compositionality}. In \bibinfo{booktitle}{\emph{Advances in neural
  information processing systems}}. \bibinfo{pages}{3111--3119}.
\newblock


\bibitem[\protect\citeauthoryear{Naumov, Mudigere, Shi, Huang, Sundaraman,
  Park, Wang, Gupta, Wu, Azzolini, et~al\mbox{.}}{Naumov et~al\mbox{.}}{2019}]%
        {DLRM}
\bibfield{author}{\bibinfo{person}{Maxim Naumov}, \bibinfo{person}{Dheevatsa
  Mudigere}, \bibinfo{person}{Hao-Jun~Michael Shi}, \bibinfo{person}{Jianyu
  Huang}, \bibinfo{person}{Narayanan Sundaraman}, \bibinfo{person}{Jongsoo
  Park}, \bibinfo{person}{Xiaodong Wang}, \bibinfo{person}{Udit Gupta},
  \bibinfo{person}{Carole-Jean Wu}, \bibinfo{person}{Alisson~G Azzolini},
  {et~al\mbox{.}}} \bibinfo{year}{2019}\natexlab{}.
\newblock \showarticletitle{Deep Learning Recommendation Model for
  Personalization and Recommendation Systems}.
\newblock \bibinfo{journal}{\emph{arXiv preprint arXiv:1906.00091}}
  (\bibinfo{year}{2019}).
\newblock


\bibitem[\protect\citeauthoryear{Qin and Zhu}{Qin and Zhu}{2013}]%
        {entropy}
\bibfield{author}{\bibinfo{person}{Lijing Qin} {and} \bibinfo{person}{Xiaoyan
  Zhu}.} \bibinfo{year}{2013}\natexlab{}.
\newblock \showarticletitle{Promoting diversity in recommendation by entropy
  regularizer}. In \bibinfo{booktitle}{\emph{Twenty-Third International Joint
  Conference on Artificial Intelligence}}.
\newblock


\bibitem[\protect\citeauthoryear{Radford, Metz, and Chintala}{Radford
  et~al\mbox{.}}{2015}]%
        {dcgan}
\bibfield{author}{\bibinfo{person}{Alec Radford}, \bibinfo{person}{Luke Metz},
  {and} \bibinfo{person}{Soumith Chintala}.} \bibinfo{year}{2015}\natexlab{}.
\newblock \showarticletitle{Unsupervised representation learning with deep
  convolutional generative adversarial networks}.
\newblock \bibinfo{journal}{\emph{arXiv preprint arXiv:1511.06434}}
  (\bibinfo{year}{2015}).
\newblock


\bibitem[\protect\citeauthoryear{Reddi, Kale, and Kumar}{Reddi
  et~al\mbox{.}}{2019}]%
        {reddi2019convergence}
\bibfield{author}{\bibinfo{person}{Sashank~J Reddi}, \bibinfo{person}{Satyen
  Kale}, {and} \bibinfo{person}{Sanjiv Kumar}.}
  \bibinfo{year}{2019}\natexlab{}.
\newblock \showarticletitle{On the convergence of adam and beyond}.
\newblock \bibinfo{journal}{\emph{arXiv preprint arXiv:1904.09237}}
  (\bibinfo{year}{2019}).
\newblock


\bibitem[\protect\citeauthoryear{Rendle}{Rendle}{2010}]%
        {FM}
\bibfield{author}{\bibinfo{person}{Steffen Rendle}.}
  \bibinfo{year}{2010}\natexlab{}.
\newblock \showarticletitle{Factorization machines}. In
  \bibinfo{booktitle}{\emph{2010 IEEE International Conference on Data
  Mining}}. IEEE, \bibinfo{pages}{995--1000}.
\newblock


\bibitem[\protect\citeauthoryear{Rendle, Freudenthaler, Gantner, and
  Schmidt-Thieme}{Rendle et~al\mbox{.}}{2009}]%
        {rendle2009bpr}
\bibfield{author}{\bibinfo{person}{Steffen Rendle}, \bibinfo{person}{Christoph
  Freudenthaler}, \bibinfo{person}{Zeno Gantner}, {and} \bibinfo{person}{Lars
  Schmidt-Thieme}.} \bibinfo{year}{2009}\natexlab{}.
\newblock \showarticletitle{BPR: Bayesian personalized ranking from implicit
  feedback}. In \bibinfo{booktitle}{\emph{Proceedings of the twenty-fifth
  conference on uncertainty in artificial intelligence}}. AUAI Press,
  \bibinfo{pages}{452--461}.
\newblock


\bibitem[\protect\citeauthoryear{Sarwar, Karypis, Konstan, Riedl,
  et~al\mbox{.}}{Sarwar et~al\mbox{.}}{2001}]%
        {cf}
\bibfield{author}{\bibinfo{person}{Badrul~Munir Sarwar},
  \bibinfo{person}{George Karypis}, \bibinfo{person}{Joseph~A Konstan},
  \bibinfo{person}{John Riedl}, {et~al\mbox{.}}}
  \bibinfo{year}{2001}\natexlab{}.
\newblock \showarticletitle{Item-based collaborative filtering recommendation
  algorithms.}
\newblock \bibinfo{journal}{\emph{Www}}  \bibinfo{volume}{1}
  (\bibinfo{year}{2001}), \bibinfo{pages}{285--295}.
\newblock


\bibitem[\protect\citeauthoryear{Sculley, Holt, Golovin, Davydov, Phillips,
  Ebner, Chaudhary, Young, Crespo, and Dennison}{Sculley et~al\mbox{.}}{2015}]%
        {sculley2015hidden}
\bibfield{author}{\bibinfo{person}{David Sculley}, \bibinfo{person}{Gary Holt},
  \bibinfo{person}{Daniel Golovin}, \bibinfo{person}{Eugene Davydov},
  \bibinfo{person}{Todd Phillips}, \bibinfo{person}{Dietmar Ebner},
  \bibinfo{person}{Vinay Chaudhary}, \bibinfo{person}{Michael Young},
  \bibinfo{person}{Jean-Francois Crespo}, {and} \bibinfo{person}{Dan
  Dennison}.} \bibinfo{year}{2015}\natexlab{}.
\newblock \showarticletitle{Hidden technical debt in machine learning systems}.
  In \bibinfo{booktitle}{\emph{Advances in neural information processing
  systems}}. \bibinfo{pages}{2503--2511}.
\newblock


\bibitem[\protect\citeauthoryear{Sha, Wu, and Niu}{Sha et~al\mbox{.}}{2016}]%
        {pmf_alpha_beta}
\bibfield{author}{\bibinfo{person}{Chaofeng Sha}, \bibinfo{person}{Xiaowei Wu},
  {and} \bibinfo{person}{Junyu Niu}.} \bibinfo{year}{2016}\natexlab{}.
\newblock \showarticletitle{A Framework for Recommending Relevant and Diverse
  Items.}. In \bibinfo{booktitle}{\emph{IJCAI}}. \bibinfo{pages}{3868--3874}.
\newblock


\bibitem[\protect\citeauthoryear{Srivastava, Hinton, Krizhevsky, Sutskever, and
  Salakhutdinov}{Srivastava et~al\mbox{.}}{2014}]%
        {dropout}
\bibfield{author}{\bibinfo{person}{Nitish Srivastava},
  \bibinfo{person}{Geoffrey Hinton}, \bibinfo{person}{Alex Krizhevsky},
  \bibinfo{person}{Ilya Sutskever}, {and} \bibinfo{person}{Ruslan
  Salakhutdinov}.} \bibinfo{year}{2014}\natexlab{}.
\newblock \showarticletitle{Dropout: a simple way to prevent neural networks
  from overfitting}.
\newblock \bibinfo{journal}{\emph{The journal of machine learning research}}
  \bibinfo{volume}{15}, \bibinfo{number}{1} (\bibinfo{year}{2014}),
  \bibinfo{pages}{1929--1958}.
\newblock


\bibitem[\protect\citeauthoryear{Szpektor, Maarek, and Pelleg}{Szpektor
  et~al\mbox{.}}{2013}]%
        {szpektor2013relevance}
\bibfield{author}{\bibinfo{person}{Idan Szpektor}, \bibinfo{person}{Yoelle
  Maarek}, {and} \bibinfo{person}{Dan Pelleg}.}
  \bibinfo{year}{2013}\natexlab{}.
\newblock \showarticletitle{When relevance is not enough: promoting diversity
  and freshness in personalized question recommendation}. In
  \bibinfo{booktitle}{\emph{Proceedings of the 22nd international conference on
  World Wide Web}}. ACM, \bibinfo{pages}{1249--1260}.
\newblock


\bibitem[\protect\citeauthoryear{Tsochantaridis, Joachims, Hofmann, and
  Altun}{Tsochantaridis et~al\mbox{.}}{2005}]%
        {tsochantaridis2005large}
\bibfield{author}{\bibinfo{person}{Ioannis Tsochantaridis},
  \bibinfo{person}{Thorsten Joachims}, \bibinfo{person}{Thomas Hofmann}, {and}
  \bibinfo{person}{Yasemin Altun}.} \bibinfo{year}{2005}\natexlab{}.
\newblock \showarticletitle{Large margin methods for structured and
  interdependent output variables}.
\newblock \bibinfo{journal}{\emph{Journal of machine learning research}}
  \bibinfo{volume}{6}, \bibinfo{number}{Sep} (\bibinfo{year}{2005}),
  \bibinfo{pages}{1453--1484}.
\newblock


\bibitem[\protect\citeauthoryear{Wang, Zhao, Xie, Li, and Guo}{Wang
  et~al\mbox{.}}{2019c}]%
        {wang2019knowledge}
\bibfield{author}{\bibinfo{person}{Hongwei Wang}, \bibinfo{person}{Miao Zhao},
  \bibinfo{person}{Xing Xie}, \bibinfo{person}{Wenjie Li}, {and}
  \bibinfo{person}{Minyi Guo}.} \bibinfo{year}{2019}\natexlab{c}.
\newblock \showarticletitle{Knowledge graph convolutional networks for
  recommender systems}. In \bibinfo{booktitle}{\emph{The World Wide Web
  Conference}}. ACM, \bibinfo{pages}{3307--3313}.
\newblock


\bibitem[\protect\citeauthoryear{Wang, He, Cao, Liu, and Chua}{Wang
  et~al\mbox{.}}{2019a}]%
        {KGAT}
\bibfield{author}{\bibinfo{person}{Xiang Wang}, \bibinfo{person}{Xiangnan He},
  \bibinfo{person}{Yixin Cao}, \bibinfo{person}{Meng Liu}, {and}
  \bibinfo{person}{Tat{-}Seng Chua}.} \bibinfo{year}{2019}\natexlab{a}.
\newblock \showarticletitle{{KGAT:} Knowledge Graph Attention Network for
  Recommendation}. In \bibinfo{booktitle}{\emph{{KDD}}}.
  \bibinfo{pages}{950--958}.
\newblock


\bibitem[\protect\citeauthoryear{Wang, He, Wang, Feng, and Chua}{Wang
  et~al\mbox{.}}{2019b}]%
        {NGCF}
\bibfield{author}{\bibinfo{person}{Xiang Wang}, \bibinfo{person}{Xiangnan He},
  \bibinfo{person}{Meng Wang}, \bibinfo{person}{Fuli Feng}, {and}
  \bibinfo{person}{Tat{-}Seng Chua}.} \bibinfo{year}{2019}\natexlab{b}.
\newblock \showarticletitle{Neural Graph Collaborative Filtering}. In
  \bibinfo{booktitle}{\emph{Proceedings of the 42nd International {ACM} {SIGIR}
  Conference on Research and Development in Information Retrieval, {SIGIR}
  2019, Paris, France, July 21-25, 2019.}} \bibinfo{pages}{165--174}.
\newblock


\bibitem[\protect\citeauthoryear{Warlop, Mary, and Gartrell}{Warlop
  et~al\mbox{.}}{2019}]%
        {warlop2019tensorized}
\bibfield{author}{\bibinfo{person}{Romain Warlop},
  \bibinfo{person}{J{\'e}r{\'e}mie Mary}, {and} \bibinfo{person}{Mike
  Gartrell}.} \bibinfo{year}{2019}\natexlab{}.
\newblock \showarticletitle{Tensorized Determinantal Point Processes for
  Recommendation}. In \bibinfo{booktitle}{\emph{Proceedings of the 25th ACM
  SIGKDD International Conference on Knowledge Discovery \& Data Mining}}. ACM,
  \bibinfo{pages}{1605--1615}.
\newblock


\bibitem[\protect\citeauthoryear{Wilhelm, Ramanathan, Bonomo, Jain, Chi, and
  Gillenwater}{Wilhelm et~al\mbox{.}}{2018}]%
        {wilhelm2018practical}
\bibfield{author}{\bibinfo{person}{Mark Wilhelm}, \bibinfo{person}{Ajith
  Ramanathan}, \bibinfo{person}{Alexander Bonomo}, \bibinfo{person}{Sagar
  Jain}, \bibinfo{person}{Ed~H Chi}, {and} \bibinfo{person}{Jennifer
  Gillenwater}.} \bibinfo{year}{2018}\natexlab{}.
\newblock \showarticletitle{Practical diversified recommendations on youtube
  with determinantal point processes}. In \bibinfo{booktitle}{\emph{Proceedings
  of the 27th ACM International Conference on Information and Knowledge
  Management}}. ACM, \bibinfo{pages}{2165--2173}.
\newblock


\bibitem[\protect\citeauthoryear{Wu, Wu, An, Huang, Huang, and Xie}{Wu
  et~al\mbox{.}}{2019d}]%
        {wu2019npa}
\bibfield{author}{\bibinfo{person}{Chuhan Wu}, \bibinfo{person}{Fangzhao Wu},
  \bibinfo{person}{Mingxiao An}, \bibinfo{person}{Jianqiang Huang},
  \bibinfo{person}{Yongfeng Huang}, {and} \bibinfo{person}{Xing Xie}.}
  \bibinfo{year}{2019}\natexlab{d}.
\newblock \showarticletitle{Npa: Neural news recommendation with personalized
  attention}. In \bibinfo{booktitle}{\emph{Proceedings of the 25th ACM SIGKDD
  International Conference on Knowledge Discovery \& Data Mining}}. ACM,
  \bibinfo{pages}{2576--2584}.
\newblock


\bibitem[\protect\citeauthoryear{Wu, Wu, An, Huang, and Xie}{Wu
  et~al\mbox{.}}{2019c}]%
        {wu2019neural}
\bibfield{author}{\bibinfo{person}{Chuhan Wu}, \bibinfo{person}{Fangzhao Wu},
  \bibinfo{person}{Mingxiao An}, \bibinfo{person}{Yongfeng Huang}, {and}
  \bibinfo{person}{Xing Xie}.} \bibinfo{year}{2019}\natexlab{c}.
\newblock \showarticletitle{Neural News Recommendation with Topic-Aware News
  Representation}. In \bibinfo{booktitle}{\emph{Proceedings of the 57th Annual
  Meeting of the Association for Computational Linguistics}}.
  \bibinfo{pages}{1154--1159}.
\newblock


\bibitem[\protect\citeauthoryear{Wu, Zhang, Souza~Jr, Fifty, Yu, and
  Weinberger}{Wu et~al\mbox{.}}{2019e}]%
        {sgc}
\bibfield{author}{\bibinfo{person}{Felix Wu}, \bibinfo{person}{Tianyi Zhang},
  \bibinfo{person}{Amauri Holanda~de Souza~Jr}, \bibinfo{person}{Christopher
  Fifty}, \bibinfo{person}{Tao Yu}, {and} \bibinfo{person}{Kilian~Q
  Weinberger}.} \bibinfo{year}{2019}\natexlab{e}.
\newblock \showarticletitle{Simplifying graph convolutional networks}.
\newblock \bibinfo{journal}{\emph{arXiv preprint arXiv:1902.07153}}
  (\bibinfo{year}{2019}).
\newblock


\bibitem[\protect\citeauthoryear{Wu, Liu, Miao, Zhao, Guan, and Tang}{Wu
  et~al\mbox{.}}{2019a}]%
        {recent}
\bibfield{author}{\bibinfo{person}{Qiong Wu}, \bibinfo{person}{Yong Liu},
  \bibinfo{person}{Chunyan Miao}, \bibinfo{person}{Yin Zhao},
  \bibinfo{person}{Lu Guan}, {and} \bibinfo{person}{Haihong Tang}.}
  \bibinfo{year}{2019}\natexlab{a}.
\newblock \showarticletitle{Recent Advances in Diversified Recommendation}.
\newblock \bibinfo{journal}{\emph{arXiv preprint arXiv:1905.06589}}
  (\bibinfo{year}{2019}).
\newblock


\bibitem[\protect\citeauthoryear{Wu, Tang, Zhu, Wang, Xie, and Tan}{Wu
  et~al\mbox{.}}{2019b}]%
        {wu2019session}
\bibfield{author}{\bibinfo{person}{Shu Wu}, \bibinfo{person}{Yuyuan Tang},
  \bibinfo{person}{Yanqiao Zhu}, \bibinfo{person}{Liang Wang},
  \bibinfo{person}{Xing Xie}, {and} \bibinfo{person}{Tieniu Tan}.}
  \bibinfo{year}{2019}\natexlab{b}.
\newblock \showarticletitle{Session-based recommendation with graph neural
  networks}. In \bibinfo{booktitle}{\emph{Proceedings of the AAAI Conference on
  Artificial Intelligence}}, Vol.~\bibinfo{volume}{33}.
  \bibinfo{pages}{346--353}.
\newblock


\bibitem[\protect\citeauthoryear{Xu, Hu, Leskovec, and Jegelka}{Xu
  et~al\mbox{.}}{2018}]%
        {powerful}
\bibfield{author}{\bibinfo{person}{Keyulu Xu}, \bibinfo{person}{Weihua Hu},
  \bibinfo{person}{Jure Leskovec}, {and} \bibinfo{person}{Stefanie Jegelka}.}
  \bibinfo{year}{2018}\natexlab{}.
\newblock \showarticletitle{How powerful are graph neural networks?}
\newblock \bibinfo{journal}{\emph{arXiv preprint arXiv:1810.00826}}
  (\bibinfo{year}{2018}).
\newblock


\bibitem[\protect\citeauthoryear{Ying, He, Chen, Eksombatchai, Hamilton, and
  Leskovec}{Ying et~al\mbox{.}}{2018}]%
        {pinsage}
\bibfield{author}{\bibinfo{person}{Rex Ying}, \bibinfo{person}{Ruining He},
  \bibinfo{person}{Kaifeng Chen}, \bibinfo{person}{Pong Eksombatchai},
  \bibinfo{person}{William~L Hamilton}, {and} \bibinfo{person}{Jure Leskovec}.}
  \bibinfo{year}{2018}\natexlab{}.
\newblock \showarticletitle{Graph convolutional neural networks for web-scale
  recommender systems}. In \bibinfo{booktitle}{\emph{Proceedings of the 24th
  ACM SIGKDD International Conference on Knowledge Discovery \& Data Mining}}.
  ACM, \bibinfo{pages}{974--983}.
\newblock


\bibitem[\protect\citeauthoryear{Zheng, Gao, He, Li, and Jin}{Zheng
  et~al\mbox{.}}{2020}]%
        {zheng2020price}
\bibfield{author}{\bibinfo{person}{Yu Zheng}, \bibinfo{person}{Chen Gao},
  \bibinfo{person}{Xiangnan He}, \bibinfo{person}{Yong Li}, {and}
  \bibinfo{person}{Depeng Jin}.} \bibinfo{year}{2020}\natexlab{}.
\newblock \showarticletitle{Price-aware recommendation with graph convolutional
  networks}. In \bibinfo{booktitle}{\emph{2020 IEEE 36th International
  Conference on Data Engineering (ICDE)}}. IEEE, \bibinfo{pages}{133--144}.
\newblock


\bibitem[\protect\citeauthoryear{Zhou, Zhu, Song, Fan, Zhu, Ma, Yan, Jin, Li,
  and Gai}{Zhou et~al\mbox{.}}{2018}]%
        {din}
\bibfield{author}{\bibinfo{person}{Guorui Zhou}, \bibinfo{person}{Xiaoqiang
  Zhu}, \bibinfo{person}{Chenru Song}, \bibinfo{person}{Ying Fan},
  \bibinfo{person}{Han Zhu}, \bibinfo{person}{Xiao Ma},
  \bibinfo{person}{Yanghui Yan}, \bibinfo{person}{Junqi Jin},
  \bibinfo{person}{Han Li}, {and} \bibinfo{person}{Kun Gai}.}
  \bibinfo{year}{2018}\natexlab{}.
\newblock \showarticletitle{Deep interest network for click-through rate
  prediction}. In \bibinfo{booktitle}{\emph{Proceedings of the 24th ACM SIGKDD
  International Conference on Knowledge Discovery \& Data Mining}}. ACM,
  \bibinfo{pages}{1059--1068}.
\newblock


\bibitem[\protect\citeauthoryear{Zhou, Kuscsik, Liu, Medo, Wakeling, and
  Zhang}{Zhou et~al\mbox{.}}{2010}]%
        {solving}
\bibfield{author}{\bibinfo{person}{Tao Zhou}, \bibinfo{person}{Zolt{\'a}n
  Kuscsik}, \bibinfo{person}{Jian-Guo Liu}, \bibinfo{person}{Mat{\'u}{\v{s}}
  Medo}, \bibinfo{person}{Joseph~Rushton Wakeling}, {and}
  \bibinfo{person}{Yi-Cheng Zhang}.} \bibinfo{year}{2010}\natexlab{}.
\newblock \showarticletitle{Solving the apparent diversity-accuracy dilemma of
  recommender systems}.
\newblock \bibinfo{journal}{\emph{Proceedings of the National Academy of
  Sciences}} \bibinfo{volume}{107}, \bibinfo{number}{10}
  (\bibinfo{year}{2010}), \bibinfo{pages}{4511--4515}.
\newblock


\bibitem[\protect\citeauthoryear{Zhu, Chang, Xu, Zhang, Li, He, Li, Xu, and
  Gai}{Zhu et~al\mbox{.}}{2019}]%
        {zhu2019joint}
\bibfield{author}{\bibinfo{person}{Han Zhu}, \bibinfo{person}{Daqing Chang},
  \bibinfo{person}{Ziru Xu}, \bibinfo{person}{Pengye Zhang},
  \bibinfo{person}{Xiang Li}, \bibinfo{person}{Jie He}, \bibinfo{person}{Han
  Li}, \bibinfo{person}{Jian Xu}, {and} \bibinfo{person}{Kun Gai}.}
  \bibinfo{year}{2019}\natexlab{}.
\newblock \showarticletitle{Joint Optimization of Tree-based Index and Deep
  Model for Recommender Systems}.
\newblock \bibinfo{journal}{\emph{arXiv preprint arXiv:1902.07565}}
  (\bibinfo{year}{2019}).
\newblock


\bibitem[\protect\citeauthoryear{Zhu, Li, Zhang, Li, He, Li, and Gai}{Zhu
  et~al\mbox{.}}{2018}]%
        {zhu2018learning}
\bibfield{author}{\bibinfo{person}{Han Zhu}, \bibinfo{person}{Xiang Li},
  \bibinfo{person}{Pengye Zhang}, \bibinfo{person}{Guozheng Li},
  \bibinfo{person}{Jie He}, \bibinfo{person}{Han Li}, {and}
  \bibinfo{person}{Kun Gai}.} \bibinfo{year}{2018}\natexlab{}.
\newblock \showarticletitle{Learning Tree-based Deep Model for Recommender
  Systems}. In \bibinfo{booktitle}{\emph{Proceedings of the 24th ACM SIGKDD
  International Conference on Knowledge Discovery \& Data Mining}}. ACM,
  \bibinfo{pages}{1079--1088}.
\newblock


\bibitem[\protect\citeauthoryear{Ziegler, McNee, Konstan, and Lausen}{Ziegler
  et~al\mbox{.}}{2005}]%
        {topic}
\bibfield{author}{\bibinfo{person}{Cai-Nicolas Ziegler},
  \bibinfo{person}{Sean~M McNee}, \bibinfo{person}{Joseph~A Konstan}, {and}
  \bibinfo{person}{Georg Lausen}.} \bibinfo{year}{2005}\natexlab{}.
\newblock \showarticletitle{Improving recommendation lists through topic
  diversification}. In \bibinfo{booktitle}{\emph{Proceedings of the 14th
  international conference on World Wide Web}}. ACM, \bibinfo{pages}{22--32}.
\newblock


\end{thebibliography}

\end{document}